\documentclass[11pt,a4paper]{article}

\usepackage{jheppub}

\usepackage{amsmath}
\usepackage{verbatim}
\usepackage{amssymb}
\usepackage{inputenc}
\usepackage{textcomp}
\usepackage{appendix}

\newcommand{\e}{\epsilon}
\newcommand{\be}[1]{\begin{equation}\label{#1} }
\newcommand{\ee}{\end{equation}}
\newcommand{\bea}[1]{\begin{eqnarray}\label{#1} }
\newcommand{\eea}{\end{eqnarray}}
\newcommand{\p}{\partial}
\newcommand{\refb}[1]{(\ref{#1})}

\renewcommand{\L}{{\mathcal{L}}}
\newcommand{\T}{{\mathcal{T}}}
\newcommand{\bL}{\bar{{\mathcal{L}}}}
\newcommand{\z}{{\bar z}}
\newcommand{\h}{{\bar h}}

\renewcommand{\>}{\rangle}
\newcommand{\<}{\langle}
\newcommand{\w}{\omega}
\newcommand{\bw}{\bar{\omega}}

\renewcommand{\a}{\alpha}
\newcommand{\ta}{\tilde{\alpha}}
\newcommand{\A}{\tilde{A}}
\newcommand{\B}{\tilde{B}}
\newcommand{\C}{\tilde{C}}
\renewcommand{\b}{\beta}
\renewcommand{\t}{\tau}
\newcommand{\s}{\sigma}

\title{Tensionless Strings from Worldsheet Symmetries}

\author[a, 1]{Arjun Bagchi} \author[b, c]{Shankhadeep Chakrabortty} \author[b, c]{and Pulastya Parekh}

\affiliation[a]{Center for Theoretical Physics, Massachusetts Institute of Technology,\\ 77 Massachusetts Avenue, Cambridge, MA 02139, USA.\\} 

\note[1]{On leave of absence from IISER Pune.} 

\affiliation[b]{Indian Institute of Science Education and Research\\ Dr Homi Bhabha Road, Pashan. Pune 411008. INDIA.\\} 

\affiliation[c]{Van Swinderen Institute for Particle Physics and Gravity, University of Groningen, \\ Nijenborgh 4, 9747 AG Groningen, The Netherlands.\\ }

\emailAdd{abagchi@mit.edu, shankhadeep.chakrabortty@iiserpune.ac.in, \\ pulastya.parekh@students.iiserpune.ac.in}

\abstract{We revisit the construction of the tensionless limit of closed bosonic string theory in the covariant formulation in the light of Galilean conformal symmetry that rises as the residual gauge symmetry on the tensionless worldsheet. We relate the analysis of the fundamentally tensionless theory to the tensionless limit that is viewed as a contraction of worldsheet coordinates. Analysis of the quantum regime uncovers interesting physics. The degrees of freedom that appear in the tensionless string are fundamentally different from the usual string states. Through a Bogoliubov transformation on the worldsheet, we link the tensionless vacuum to the usual tensile vacuum. As an application, we show that our analysis can be used to understand physics of strings at very high temperatures and propose that these new degrees of freedom are naturally connected with the long-string picture of the Hagedorn phase of free string theory. We also show that tensionless closed strings behave like open strings.}

\preprint{MIT-CTP-4690}

\begin{document}
\maketitle

\section{Introduction}

String theory is, at present, the best framework for understanding a theory of quantum gravity. This generalises the framework of quantum field theory, which is based on point-particles, to fundamental one-dimensional strings and, using quantum mechanics and the special theory of relativity as inputs, naturally generates a quantum theory of general relativity. One of the principal reasons that we are successful in understanding the theory of quantised strings is the power of symmetry, especially conformal symmetry which arises on the worldsheet of the string as a residual gauge symmetry in the conformal gauge. The lack of conformal symmetry is also the principal reason why attempts at understanding a theory of quantised higher dimensional membranes has eluded us thus far. 

\subsection{High energy string theory}

The study of the extreme high energy limit of string theory has remained an enigmatic subject since the discovery by Gross and Mende that string scattering amplitudes behave in a particularly simple way in this limit \cite{Gross:1987kza, Gross:1987ar}. String theory is a very unique framework in the sense that it is described by only a very few parameters. It is thus very likely that it would have a very high degree of symmetry. This is however not manifest in the formulation of the theory. The likely explanation for this lies in the analogous field theory example. If one considers a spontaneously broken gauge theory, the low energy dynamics of the theory would never capture the details of the unbroken gauge group. But it is likely, just as in the case of the electro-weak theory in a limit where the gauge boson masses are negligible, that the high energy scattering amplitudes see the unbroken larger symmetry structure. Extending this analogy to string theory where the analogue of the weak-scale would be the Planck mass, in \cite{Gross:1987kza, Gross:1987ar}, the authors explored a particular limit of string theory.  This limit, where the Planck mass would be negligible, phrased in terms of the only string variable $\alpha'$, was the $\alpha' \to \infty$ or the tensionless sector  of string theory. This limit is the diametric opposite to the limit to supergravity where one takes $\a' \to 0$ making the string shrink to a point. The tensionless sector of strings is expected to capture a large hitherto unknown symmetry of string theory. In \cite{Gross:1988ue}, Gross found an infinite number of linear relations between the scattering amplitudes of different string states that are valid order by order in perturbation theory as the string tension went to zero. This was an indication of the aforesaid symmetry structure. 

The initiation of the study of tensionless or null strings predates the above discussions by a decade and the first effort in this goes back to \cite{Schild}. In a modern perspective, the study of tensionless strings has been linked with the emergence of higher spin symmetries in the space-time. Vasiliev's higher spin theories \cite{Vasiliev:2004qz} hold the promise for realizing at least partially the huge unseen symmetry of string theory described above. The tensionless limit of string theory is the limit where the masses of all states go to zero and hence one obtains massless higher spin states in this sector \cite{Witten-talk, Sundborg:2000wp}. The hope is that the generation of masses in string theory would occur by the breaking of this higher spin theory. All of this has come to the forefront of research in the recent years with the discovery of new higher spin holographic dualities following Klebanov-Polyakov-Sezgin-Sundell \cite{Klebanov:2002ja,Sezgin:2002rt} and more recently Gaberdiel-Gopakumar \cite{Gaberdiel:2010pz}, and subsequent attempts to connect these to string theory (e.g. see \cite{Chang:2012kt} -- \cite{Gaberdiel:2015uca}). 

\subsection{Symmetries of tensionless strings}

Our motivations in the current paper are somewhat more old-fashioned. We consider the theory of tensionless strings propagating in flat space and wish to construct the theory from first principles in the analogue of the conformal gauge following \cite{Isberg:1993av}. The reason why we wish to do this is because like in the case of tensile strings, where conformal field theory offers a guiding principle for the construction of the theory, it has been established that a new symmetry structure governs the theory of tensionless strings. This is the so-called Galilean conformal symmetry, which has made its appearance earlier in the context of the non-relativistic limit of AdS/CFT \cite{Bagchi:2009my} and somewhat more surprisingly in the construction of flat holography \cite{Bagchi:2010eg}. Galilean conformal symmetry arises in the case of the tensionless string as the residual symmetry on the worldsheet when the analogue of the conformal gauge is fixed \cite{Bagchi:2013bga}. We wish to make use of the techniques of Galilean Conformal Field Theories (GCFTs) which have been developed earlier, now to the case of the tensionless string and thereby provide a organizing principle to the study of the subject. 

Like the early work in the usual tensile case, we begin by considering the theory of tensionless strings on a flat background. This may be an immediate concern for the reader. The tensionless limit of string theory, as described before, is connected with the theory of higher spins and thus there is an obvious problem. In flat space times of dimensions four and higher, there is no consistent theory of interacting higher spins. As shown by Fradkin and Vasiliev \cite{Fradkin:1987ks}, consistent interacting higher spin theories require a non-zero cosmological constant. This would mean that once we include interactions in our theory, we would inevitably be lead to inconsistencies. In spite of the above, there are many reasons that we want to consider tensionless strings in flat space times and let us list these now. First, in this paper, we are interested only in free tensionless strings and thus would not encounter these problems. Secondly, we want to set up the formalism of tensionless strings in the light of the Galilean conformal symmetry on the worldsheet and exploit the known symmetry structure to understand this rather old problem. Thus understanding the simplest case of the strings in the flat background is the obvious first task before one attempts to understand it in more complicated situations like AdS space times which this programme eventually wishes to address. Lastly, we want to mention that there is a theory of higher spins in flat spacetimes that is free from the above difficulties and this is a theory in three dimensions \cite{Afshar:2013vka, Gonzalez:2013oaa}. Our present considerations may be of use there. It is interesting to note that in light of recent work in strings in three space-time dimensions which give rise to anyonic particles in its spectrum \cite{Mezincescu:2010yp}, there has been an effort to understand the tensionless limit of this theory \cite{Murase:2013via}. Since three dimensions is usually simpler to deal with, it would be of interest to see how all of these fit together with our analysis in this paper, adapted for the $D=3$ case, fits together with the analysis of \cite{Mezincescu:2010yp, Murase:2013via}. A list of other older work on the subject most relevant for our present paper is \cite{Karlhede:1986wb} -- \cite{Francia:2007qt}. 

\subsection{New degrees of freedom: a window to Hagedorn physics?}
In the present paper, as described above, we look at the construction of the tensionless string from the point of the symmetries of the worldsheet, both in terms of a limiting principal as well as a fundamental analysis. A brief description of the outline is provided later, but let us stress on one particularly important point. Perhaps the most interesting finding of this paper is the natural emergence of new degrees of freedom in the tensionless limit. As will be made clear in our analysis, these new degrees of freedom are very different from the usual perturbative states. We discover a new vacuum which we call the tensionless vacuum that is clearly not the same as the usual vacuum of perturbative tensile string theory. The new vacuum and the excitations around it are described in terms of new creation and annihilation operators which can be related to the usual ones by Bogoliubov transformations on the worldsheet. The new tensionless vacuum can be understood as a coherent or squeezed state in terms of the oscillators of the tensile theory. We make these mappings precise in our work and offer an explanation of how this new worldsheet phenomenon can be linked to physics of strings at very  high temperatures. 

The extreme high energy limit of string theory has been of interest due to the possible existence of a mysterious Hagedorn phase \cite{Atick:1988si}. The statistical mechanics of string theory at high temperatures is very unlike the behaviour of point-particles. Here we encounter an exponential growth in the single-string density of states. The string theory partition function converges only at temperatures less than a certain limiting temperature $T_H$ known as the Hagedorn temperature. This has been given different physical interpretations in the literature. One suggestion is that the Hagedorn temperature defines an absolute limiting temperature. Another interpretation is that this signals a transition to a phase where the fundamental degrees of freedom differ from usual string theory. 

When considering the non-interacting theory, it can be shown that as one approaches the Hagedorn temperature, it becomes thermodynamically more favourable to form a single long string as opposed to heating up a gas of strings \cite{Bowick:1989us, Giddings:1989xe}. This picture is made more complicated when one includes interactions.  

It has been long suspected that the tensionless limit of string theory would have something to say about the physics at the Hagedorn scale as the effective string tension at $T_H$ goes to zero. We interpret the natural emergence of the aforementioned new degrees of freedom in the tensionless limit as the worldsheet manifestation of the onset of the Hagedorn phase. The tensionless vacuum, we claim, is the emergent long string at the Hagedorn temperature. As opposed to the popular view that the worldsheet picture breaks down near the Hagedorn temperature, we advocate that there indeed is a worldsheet, albeit very different from the usual tensile case. We also suggest that the thermal nature of this vacuum can be understood by entanglement between the left and right moving degrees of freedom on the original tensile string theory worldsheet.

\subsection{Plan of the paper}
Having motivated our goals in the present paper, let us now go on to describe the contents and structure of our work. 

We begin by reviewing the symmetries of the tensionless closed bosonic string following \cite{Isberg:1993av} in Sec.~\ref{S2}. Parts of our paper is motivated by this work and we shall try to elaborate and extend their analysis. The connection to Galilean conformal symmetry was first made in \cite{Bagchi:2013bga} and it was shown how one can view the limit on the string worldsheet. We briefly review this and also enlist some of the basic features of GCFTs, especially in two dimensions, in an accompanying appendix (Appendix \ref{ApA}).  The uninitiated reader is encouraged to read this for a quick review of GCFTs. 

In Sec.~\ref{S3}, we present the oscillator construction for the solutions of the classical string. We show how the constraints nicely fit into the structure of the GCFT described in the above mentioned appendix. We recover our answers by looking at the systematic limiting procedure from the tensile string theory. We comment on another limit which was presented in \cite{Bagchi:2013bga} and conjectured to lead to the same local physics. Further details of this limit are presented in an appendix (Appendix \ref{ApB}).

We venture into the quantum aspects of the theory after this and first show how the usual lore of the link between tensionless strings and higher spins appear in our picture in Sec.~\ref{S4}. We then consider the fundamentally tensionless theory in Sec.~\ref{S5} and show the emergence of the new vacuum state. We show how worldsheet Bogoliubov transformations link the tensile and tensionless sectors. 

In Sec.~\ref{S6}, we turn our attentions to particular examples where our construction of the tensionless string is applicable. We devote the bulk of this section to connections of the tensionless strings to Hagedorn physics and also comment on other situations where tensionless strings appear. We conclude with a summary, further comments and a roadmap of possible future directions in Sec.~\ref{S7}.

\newpage

\section{Symmetries of tensionless strings}\label{S2}

In this section, we shall first review the construction of the action of the tensionless strings and its symmetries following \cite{Isberg:1993av}. Then we shall remind the reader how one can view this limit as the scaling of co-ordinates on the worldsheet following \cite{Bagchi:2013bga}. 

\subsection{Classical tensionless closed strings: constructing the action}
We start off with a description of the classical theory of tensionless strings \cite{Isberg:1993av}. The Nambu-Goto action of the usual tensile bosonic string is:
\be{}
S = \T \int d^2 \xi \sqrt{-\det \gamma_{\alpha \beta}} 
\ee
where $\xi_a$ are worldsheet coordinates, $\T$ is the tension and $\gamma_{\alpha \beta}$ is the induced metric 
\be{} 
\gamma_{\alpha \beta} = \p_\alpha X^m \p_\beta X^n \eta_{mn}.
\ee
Here $X^n$ are the spacetime co-ordinates of the string and $\eta_{mn}$ is the flat background metric. Taking the tensionless limit on this action is not possible as we have an explicit factor of the tension multiplying the action. Following \cite{Isberg:1993av}, we resort to rewriting the action in a form which makes taking the limit easier. To this end, we derive the generalised momenta: 
\be{}
P_m = \frac{\p\mathcal{L}}{\p\dot{X}^m}=\T \frac{(\dot{X}\cdot X')X'_{m}-(X'^2)\dot{X}_m}{\sqrt{(\dot{X}\cdot X')^2-(\dot{X}^2)(X'^2)}} = \T \sqrt{-\gamma} \gamma^{0 \alpha} \p_\alpha X_m
\ee
where the dot represents differentiation with respect to worldsheet time $\t$ and the dash represents that with respect to $\s$, the spatial worldsheet coordinate. The generalised momenta satisfy the following constraints
\be{}
P^2 + \T^2 \gamma \gamma^{00}=0, \,\ P_m\p_{\alpha} X^m =0.
\ee
The canonical Hamiltonian of the system vanishes due to the diffeomorphism invariance of the worldsheet. The total Hamiltonian of the system is just made up of the constraints: 
\be{}
\mathcal{H} = \lambda(P^2 + \T^2 \gamma \gamma^{00}) + \rho^\alpha P_m\p_{\alpha} X^m .
\ee
We rewrite the action and then integrate out the momenta to obtain
\be{act}
S = \frac{1}{2} \int d^2 \xi \, \frac{1}{2 \lambda} \bigg[ {\dot{X}}^2 -2 \rho^\alpha{\dot{X}}^m \p_\alpha X_m + \rho^a \rho^b \p_b X^m \p_a X_m - 4 \lambda^2 \T^2 \gamma \gamma^{00}\bigg].
\ee
To recast the action \refb{act} in the familiar Weyl invariant form
\be{weyl}
S = -\frac{\T}{2}  \int d^2 \xi \sqrt{-g} g^{\alpha \beta} \p_\alpha X^m \p_\beta X^n \eta_{mn},
\ee
we identify
\be{}
g^{\alpha \beta} = \begin{pmatrix} -1 & \rho \\ \rho & -\rho^2 + 4 \lambda^2 \T^2 \end{pmatrix}.
\ee
The tensionless limit can be taken from \refb{act} or later. It is interesting to note here that the metric density would degenerate $\T \sqrt{-g} g^{\alpha \beta}$ in the limit. One can replace this by a rank one matrix that can be written as $V^\alpha V^\beta$ where $V^\alpha$ is a vector density
\be{}
V^\alpha = \frac{1}{\sqrt{2} \lambda} (1, \rho).
\ee
The action \refb{act} in the $\T \to 0$ limit then becomes 
\be{acti}
S = \int d^2 \xi \,\ V^\alpha V^\beta \p_\alpha X^m \p_\beta X^n \eta_{mn}.
\ee

\subsection{Residual symmetries}
Under a diffeomorphism $\xi^\a\rightarrow \xi^\a+\epsilon^\a$, the vector density $V^\a$ transforms as:
\be{} \delta V^\a=-V^\beta\p_\beta \epsilon^\a + \epsilon^\beta \p_\beta V^\a +\frac{1}{2}(\p_\beta \epsilon^\beta) V^\a. \ee
The action of the tensionless string is invariant under these worldsheet diffeomorphisms and hence we need to fix a gauge.  
It is particularly useful to look at the tensionless action in the analogue of the conformal gauge for the tensile string
\be{v0}
V^\alpha = (v, 0), 
\ee
where $v$ is a constant. Just as in the tensile case, there is a residual symmetry that is left over after this gauge fixing. In the tensile case, the residual symmetry in the conformal gauge is two copies of the Virasoro algebra for the closed string. This infinite dimensional symmetry structure has been central to understand the theory of usual tensile strings. The form of $\e^\a$ which leaves the gauge fixed action (the action \refb{acti} in gauge \refb{v0}) invariant is
\be{e-ur} 
\epsilon^\a=(f'(\sigma)\tau+g(\sigma),f(\sigma)). 
\ee
For a function $F(\xi^a)$, The effect of such a transformation is given by:
\be{} \delta F=[f'(\sigma)\tau\p_\tau+f(\sigma)\p_\sigma+g(\sigma)\p_\tau]F =[L(f)+M(g)]F. \ee
Thus the generators can be defined as: 
\be{} L(f)=f'(\sigma)\tau\p_\tau+f(\sigma)\p_\sigma, \quad M(g)=g(\sigma)\p_\tau. \ee
They satisfy the commutation relations:
\be{} [L(f_1),L(f_2)]=L(f_1f'_2-f'_1f_2), \quad [L(f),M(g)]=M(fg'-f'g), \quad [M(g_1),M(g_2)]=0. \ee
The symmetry algebra can be cast in the following form:
\be{} 
[L_m,L_n]=(m-n)L_{m+n}, \quad [L_m,M_n]=(m-n)M_{m+n}, \quad [M_m,M_n]=0. 
\ee
where we can expand functions $f$, $g$ in terms of fourier modes: $f=\sum a_n e^{in\sigma},\quad g=\sum b_n e^{in\sigma}$ and written: 
\be{} 
L(f)=\sum_n a_n e^{in\sigma}(\p_\sigma+in\tau\p_\tau)=-i\sum_n a_n L_n, \quad M(g)=\sum_n b_n e^{in\sigma} \p_\tau=-i\sum_n b_n M_n.  
\ee
The algebra of the generators of the residual gauge symmetry can be identified with the 2D Galilean Conformal Algebra \refb{gca2d}. We refer the uninitiated reader to Appendix \ref{ApA} for a quick tour of the basics of the GCA and its field theoretic aspects.

\subsection{Tensionless limit from worldsheet contractions}

For the tensile string, two copies of the Virasoro algebra arise as the residual symmetry in the conformal gauge, $g_{\a \b} = e^{\phi} \eta_{\a \b}$. In case of the tensionless strings, the Virasoro symmetry is replaced by 2D Galilean conformal symmetry in the equivalent of the conformal gauge as discussed above. The form of the generators is given by
\be{genr} 
L_n = ie^{in\s}( \p_\s+ in\t \p_\t), \quad M_n =  i  e^{in\s} \p_\t. 
\ee
Again, referring the reader back to Appendix \ref{ApA}, we note that \refb{genr} is identical to the ultra-relativistic limit \refb{ur-lim} of the conformal algebra where one performs a contraction of a linear combination of the two copies of the Virasoro algebra for the tensile residual symmetries. 

The physical interpretation of this limit on the string theory worldsheet is quite straightforward and discussed at length in \cite{Bagchi:2013bga}. Here we recall the basic argument. The tensionless limit of string theory is the limit where the string becomes floppy and the length of the string becomes infinite. Re-expressed in terms of co-ordinates on the worldsheet, this limit can be interpreted as a limit of $\s \to \infty$. Since the ends of a closed string are identified $\s \sim \s + 2\pi$, instead of sending $\s \to \infty$, this is better viewed as a limit where $(\s \to \s, \t \to \e \t, \e \to 0)$. In terms of worldsheet velocities $v$ and the worldsheet speed of light $c$, this is $\frac{v}{c} = \frac{\s}{\t} \to \infty$ and hence the limit where the speed of light on the worldsheet goes to zero. The interpretation of the tensionless limit in terms of an ultra-relativistic limit on the worldsheet is thus justified. 

With the relativistic central charges $c, \bar{c}$ included in the Virasoro algebra, the limit leads to the quantum 2d GCA \refb{gca2d}. The central charges have the expression $c_L = c - \bar{c}$, $c_M= \e ( c + \bar{c} )$ \refb{c-ur}. When the original CFT has $c = \bar{c}$, then $c_L=0$, which means for parent theories without diffeomorphism anomalies the Virasoro central term of the GCFT is zero.  We also note that to keep $c_M$ finite, we would need to scale $c, \bar{c}$ to be infinitely large. This essentially means that we would have to deal with an infinite number of fields on the worldsheet. Thus, if we are interested in the tensionless sector of a well-defined string theory, it must have vanishing $c_L$ and $c_M$ \cite{Bagchi:2013bga}. This is in keeping with the various claims in the literature that the tensionless string is consistent in any spacetime dimension \cite{Lizzi:1986nv}. 

We should remark that it is possible to think of tensionless strings which are not derived as a limit from tensile string theories and are fundamental objects in their own right.  The analysis of these strings may lead to residual symmetry algebras that are the 2d GCA but with non-zero central terms as there is no algebraic constraint forcing the central terms $c_L, c_M$ to be zero.

\section{Tensionless strings: oscillator construction}\label{S3}

In this section, we look at various aspects of the tensionless string. We start out by looking at the conserved charges from the tensionless action and show that these close to form the GCA. We then move on to the analysis of the equations of motion of the tensionless string and work out its mode expansion. We show how GCFT structures naturally arise from the analysis and also recover various results by looking carefully at the limit from the tensile string theory.

\subsection{Equations of motion and tensionless mode expansions}\label{TEom}
The equations of motion of the tensionless string are easily obtainable from the tensionless action \refb{acti}. These are 
\be{eom} 
\p_\a(V^\a V^\b \p_\b X^\mu)=0, \quad V^\b\gamma_{\a \b}=0. 
\ee
The second equation in \refb{eom} indicates that the metric $\gamma_{\a \b}$ is degenerate \cite{Isberg:1993av}. In our gauge of interest, viz. $V^\a=(v,0)$, the equations take a particularly simple form:
\be{v0eom}
\ddot{X}^\mu=0; \quad \dot{X}^2=0, \quad \dot{X}\cdot X'=0. 
\ee
This implies that the tensionless string behaves as a bunch of massless point particles constrained to move transversely to the direction of the string. 

We now concentrate on the solutions of the equations of motion. For this we are interested in a mode expansion for tensionless string. A convenient expansion which solves \refb{v0eom} is the following:  
\be{mode-tless} 
X^{\mu}(\sigma,\tau)=x^{\mu}+\sqrt{2c'}A^{\mu}_0\sigma+\sqrt{2c'}B^{\mu}_0\tau+i\sqrt{2c'}\sum_{n\neq0}\frac{1}{n} \left(A^{\mu}_n-in\tau B^{\mu}_n \right)e^{-in\sigma}. 
\ee
Here $c'$ is a constant with the dimensions of $[L]^{2}$ which replaces $\a'$ that appears in the usual tensile string expansion. We demand that the tensionless closed string satisfy the boundary condition 
\be{}X^\mu(\tau,\sigma)=X^\mu(\tau,\sigma+2\pi).
\ee
In order for the mode expansion \refb{mode-tless} to be valid, we must then have have $A^{\mu}_0=0$. 
The derivatives w.r.t $\tau$ and $\sigma$ are:
\be{}
X'^{\mu}=\sqrt{2c'}\sum_{n} \left(A^{\mu}_n-in\tau B^{\mu}_n \right)e^{-in\sigma}, \quad  \dot{X}^{\mu}=\sqrt{2c'}\sum_{n} B^{\mu}_ne^{-in\sigma}. 
\ee
The two contraints of the system are $\dot{X}^2=0$ and $X'\cdot \dot{X}=0$. This translates to: 
\bea{}  
\dot{X}^2 &=&2c'\sum_{n} \left[\sum_{m} B_{-m}\cdot B_{m+n}\right]e^{-in\sigma}=\sum_{n} M_n e^{-in\sigma}=0, \label{stress1} \\ 
X'\cdot \dot{X} &=&2c'\sum_{n, m} (A_{- m}-in\tau B_{- m})\cdot B_{m+n} e^{-in\sigma}=\sum_{n} \left[L_n-in\tau M_n\right] e^{-in\sigma}=0. \label{stress2}
\eea
It is illuminating at this juncture to point out that the above expressions of the constraints translate to conditions on the GCFT stress tensor, if we compare \refb{stress1} with \refb{Tur2} and \refb{stress2} with \refb{Tur1}. Thus, the constraints of the tensionless string take the form:
\be{TT}
T_1 (\s, \t) = \sum_{n} \left[L_n-in\tau M_n\right] e^{-in\sigma} = 0, \quad T_2 (\s, \t) = \sum_{n} M_n e^{-in\sigma}=0 
\ee
in analogy with the usual tensile theory where the constraints are linked with the EM tensor of the Virasoro algebra. (Note that here we have switched off $c_L$ and $c_M$. This is because we are still in the classical regime.)

In the above, we have introduced: 
\be{} L_n= \frac{1}{2} \sum_{m} A_{- m}\cdot B_{m+n}, \quad M_n= \frac{1}{2} \sum_{m} B_{-m}\cdot B_{m+n} \ee
The classical constraints can be equivalently written as 
\be{}
L_n=0, \quad M_n = 0.
\ee
We shall later proceed to use these as constraints on the physical states on the Hilbert space in the quantum theory when we look to build a theory of quantum tensionless strings. Let us check the Poisson bracket between our newly constructed oscillator algebra. The Poisson bracket relations between X and P require
\be{AB} 
\{A^{\mu}_m,A^{\nu}_n\}_{P.B}= \{B^{\mu}_m,B^{\nu}_n\}_{P.B}=0, \quad \{A^{\mu}_m,B^{\nu}_n\}_{P.B}= - 2 im\delta_{m+n}\eta^{\mu \nu} 
\ee
Correspondingly this gives the relations between $L_m$ and $M_n$ s:
\be{} 
\{L_m,L_n\}_{P.B} =-i(m-n)L_{m+n} \quad \{L_m,M_n\}_{P.B} =-i(m-n)M_{m+n} \quad \{M_m,M_n\}_{P.B} =0 
\ee 
This is the classical version of the GCA which when quantized in the canonical way $\{ \, , \}_{P.B} \to - \frac{i}{\hbar} [\, ,]$ yields the usual GCA \refb{gca2d}.  

\subsection{Conserved charges of the tensionless string}
To lend further support to the connection of the energy-momentum tensor of the GCFT with the constraints of the tensionless string, let us now look at the construction of charges directly from the tensionless action \refb{acti}. We consider an infinitesimal transformation of \refb{acti}
\be{trans} 
\sigma^\a\rightarrow\sigma'^\a=\sigma^\a+\delta\sigma^\a  
\ee
The Noether current is given by
\be{} 
J^\a=T^\a_{\ \beta}\delta\sigma^\beta 
\ee
We can construct the energy momentum tensor by looking at the above transformation \refb{trans} on the action \refb{acti} and this turns out to be: 
\be{} 
T^\a_{\ \beta}=V^\a V^\rho \p_\rho X^\mu \p_\beta X_\mu-\frac{1}{2}V^\lambda V^\rho \p_\lambda X^\mu\p_\rho X_\mu \delta^\a_\beta 
\ee
We have been interested in the gauge choice $V^\a=(v,0)$.  The form of $\delta\sigma^\a$ in this case is:
\be{} \delta\sigma^\a=(f'\tau+g,f) \ee
where $f$ and $g$ are functions of $\sigma$ only. The non-trivial components of $T^\a_{\ \beta}$ are
\be{} 
T^0_{\ 0}=-T^1_{\ 1}=\frac{1}{2}v^2 \dot X^2, \quad T^0_{\ 1}=v^2 \dot X\cdot X' 
\ee
So, we see that 
\be{}
T^0_{\ 1} = T_1 (\s, \t), \quad T^0_{\ 0}=-T^1_{\ 1} = T_2 (\s, \t)
\ee
We integrate the zeroth component of the Noether current to generate the charge, which is given by: 
\be{}
 Q=\int d\sigma J^0 = \int d\sigma \left[T^0_{\ 0}(f'\tau+g)+T^0_{\ 1}f \right] = \int d\sigma \left[(T^0_{\ 0}f'\tau+T^0_{\ 1}f)+T^0_{\ 0}g \right]
\ee
Expanding $f$ and $g$ in fourier modes: $f=\sum a_ne^{in\sigma}$ and $g=\sum b_ne^{in\sigma}$ we get:
\be{}  
Q=\sum_n a_n \int d\sigma (T^0_{\ 0} in\tau+T^0_{\ 1})e^{in\sigma}+\sum_n b_n \int d\sigma T^0_{\ 0}e^{in\sigma} = \sum_n a_n L_n +\sum_n b_n M_n 
\ee
Thus we have:
\be{} 
L_n=\int d\sigma \left[T^0_{\ 0}in\tau+T^0_{\ 1}\right]e^{in\sigma}, \quad M_n=\int d\sigma\ T^0_{\ 0}\ e^{in\sigma} 
\ee
So in the above, we have derived the expressions for the charges from a Noetherian prescription. Needless to say, these charges close to form the 2d GCA. The point of this exercise was to show that the construction of the EM tensor and the generators of the GCA can be achieved by looking at the construction of charges as well. 

\bigskip

\subsection{Limit from tensile closed string} \label{limit-t}
The mode expansion and the oscillator algebra above were derived ``intrinsically" from the equations of motion and we did not talk about any limiting procedure in the above. It is important to check whether we can arrive at the same expressions by taking a careful limit of appropriate expressions of the tensile string theory. We start off by comparing mode expansions. The mode expansion for closed tensile string is given by 
\be{t-exp} 
X^{\mu}(\sigma,\tau)=x^{\mu}+2\sqrt{2\alpha'}\alpha^{\mu}_0\tau+i\sqrt{2\alpha'}\sum_{n\neq0}\frac{1}{n}[\a^{\mu}_ne^{-in(\tau+\sigma)}+\ta^{\mu}_ne^{-in(\tau-\sigma)}] 
\ee
where $\alpha^{\mu}_0=\tilde{\alpha}^{\mu}_0$. We know that the process of contraction to the tensionless string entails taking the limit $\tau\rightarrow\e \tau$ and $\sigma\rightarrow\sigma$. We are looking at the tensionless limit and hence $\alpha'$ should also transform as $\alpha'\rightarrow c'/\e$, where $c'$ is finite. Thus the modes for closed tensile string take the form:
\be{expal} 
X^{\mu}(\sigma,\tau) =x^{\mu}+2\sqrt{2c'}(\sqrt{\e})\alpha^{\mu}_0\tau+i\sqrt{2c'}\sum_{n\neq0}\frac{1}{n}\left[\frac{\alpha^{\mu}_n-\tilde{\alpha}^{\mu}_{- n}}{\sqrt{\e}}-in\tau\sqrt{\e}({\alpha}^{\mu}_{n}+\tilde{\alpha}^{\mu}_{- n})\right]e^{-in\sigma}
\ee
Comparing with Eqn \refb{mode-tless}, we find that
\be{ABmap} A^{\mu}_n =\frac{1}{\sqrt{\e}}({\alpha}^{\mu}_n-\tilde{\alpha}^{\mu}_{- n}), \quad
B^{\mu}_n =\sqrt{\e}({\alpha}^{\mu}_n+\tilde{\alpha}^{\mu}_{- n}) 
\ee
Using the relations above \refb{ABmap} and the Poisson bracket relations between $\alpha$'s and $\ta$'s:
\be{} \{\alpha^{\mu}_m, \alpha^{\mu}_n\}_{P.B}=-im\delta_{m+n}\eta^{{\mu\nu}}= \{\tilde{\alpha}^{\mu}_m,\tilde{\alpha}^{\mu}_n\}_{P.B}, \quad \{\alpha^{\mu}_m, \tilde{\alpha}^{\mu}_n\}_{P.B} = 0,
\ee 
we get back the relations \refb{AB} that we obtained earlier. 

It is also instructive to look at the constraints of the tensile string and take the limit and obtain the constraints in the tensionless case that we have obtained above. Due to the conspiracy of factors of $\e$, one needs to be careful in this analysis. We will be explicit in this calculation to demonstrate that one needs to be careful with these factors of $\e$. 
The constraint in the tensile case is: 
\be{} 
\dot{X}^2+X^{'2}=0 
\ee 
On taking the limit $\tau\rightarrow\e\tau$ :
\be{}
\frac{1}{\epsilon^2} \dot{X}^2+X^{'2}=0 \quad \text{or} \quad \dot{X}^2+{\epsilon^2}X^{'2}=0 
\ee %L
This is equivalent of the constraint $\dot{X}^2=0$ in the fundamental case. It is very important to keep track of this extra $\e^2$ piece in the above equation. Starting from the mode expansion \refb{expal}, one can take explicit derivatives with respect of $\t$ and $\s$ and square them to obtain
\be{33}
\dot{X}^2  = 2c'\sum_{m} \sum_{n} \e\left[{\alpha}_{- m}\cdot {\alpha}_{m+n} +  {\alpha}_{- m}\cdot \ta_{-m-n}+ \ta_{m}\cdot{\alpha}_{m+n} +  \ta_{- m} \cdot \ta_{m-n} \right]e^{-in\sigma} \ee
Taking $\e^2 X'^2$ and keeping terms upto 1st order in $\e$
\be{34} 
{\epsilon^2}X^{'2}=2c'\sum_{m} \sum_{n}\e \left[ {\alpha}_{- m}\cdot{\alpha}_{m+n}-{\alpha}_{-m}\cdot\ta_{-m-n}-\ta_{m}\cdot{\alpha}_{m+n}+\ta_{- m}\cdot\ta_{m-n} \right]e^{-in\sigma} \end{equation}
So adding the equations \refb{33} and \refb{34} above we get 
\be{}
\dot{X}^2+{\epsilon^2}X^{'2}\approx \dot{X}^2=4c'\sum_{m, n} \e\left[{\alpha}_{- m}\cdot {\alpha}_{m+n} +  \ta_{- m} \cdot \ta_{m-n} \right]e^{-in\sigma}=4c' \sum_{n} \e\left[{\L}_n +  {\bL}_{- n} \right]e^{-in\sigma}=0
\ee
Comparing with \refb{TT} we identify:
\be{} 
M_n=\e\left[{\L}_n +  \bar{\L}_{- n} \right] 
\ee
Similarly the other constraint  gives 
\be{} L_n =\sum_m ({\alpha}_{- m}\cdot {\alpha}_{m+n}-\ta_{- m}\cdot\ta_{m- n}) ={\L}_n-\bar{\L}_{- n}\ee
Thus the tensionless constraints in terms of the tensile ones in this limit generate the UR limit of the residual conformal algebra to the GCA \refb{ur-lim} on the string theory worldsheet:
\be{} 
L_n={\L}_n-\bar{\L}_{- n}, \quad M_n=\e\left[{\L}_n +  \bar{\L}_{- n} \right]. 
\ee

\bigskip

\subsection{The non-relativistic limit on the worldsheet} \label{NRws}

At this juncture, it is of interest to revisit one of the novel claims of \cite{Bagchi:2013bga}. There it was argued that if one works on the Euclidean worldsheet, there is no distinction locally between the $\s$ and $\t$ direction. Hence, a contraction of one co-ordinate would be equivalent to a contraction of the other, i.e. the contractions $(\s, \t) \to (\s, \e \t)$ and $(\s, \t) \to (\e \s, \t)$ should locally yield the same physics. This is rather counter intuitive as we are claiming that the tensionless limit of string theory, where the fundamental string becomes infinitely long, share features with the point particle limit. 

The limit on the worldsheet described above is the ultra-relativistic limit where one takes the worldsheet speed of light to zero. This, as we have seen above, is the $(\s, \t) \to (\s, \e \t)$ limit and is intimately linked to the choice of gauge \refb{v0}. 

We now will show that the other limit, viz. $(\s, \t) \to (\e \s, \t)$, which can be interpreted as a worldsheet non-relativistic limit,   arises out of a different choice of gauge in the tensionless string. We wish to look at the tensionless string theory action \refb{acti} again which has  gauge symmetry and we need to fix gauge. The gauge of choice now would be 
\be{}
V^\a = (0, v)
\ee
which is clearly different from \refb{v0}. As before, there is a residual gauge symmetry. It can easily be seen that the gauge fixed action is now left invariant under infinitesimal variations $\e^\a$ of the form
\be{e-nr} 
\epsilon^\a=\{ \s \frac{d}{d \t}{f}(\t)+g(\t),f(\t) \} 
\ee
which is just the $\s \leftrightarrow \t$ flipped version of \refb{e-ur}. The rest of the analysis follows trivially with the same exchanges of the co-ordinates and we can show that the residual symmetry is the 2D GCA now with generators 
\be{gen} 
L_n = ie^{in\t}( \p_\t+ in\s \p_\s), \quad M_n =  i  e^{in\t} \p_\s. 
\ee
We recognise this to be the generators \refb{cyl-NR} which arise from the non-relativistic contraction of the Virasoro algebra \refb{nr-lim}. Hence we see that at the level of symmetries, on the worldsheet the two different contractions are related just by a different choice of gauge. We can push further with this limit and in Appendix \ref{ApB}, we perform the mode expansions and discuss the limiting procedure which throws up some peculiarities.

\newpage

\section{Tensionless strings and higher spins}\label{S4}
In this section, we explore the much anticipated claim of a link between tensionless strings and higher spin theories in our formalism. The analysis of constraints in the previous sections has lead us to the following physical state conditions that are to be imposed on the states of the Hilbert space: 
\be{}
\< \Phi_1| T_1 |\Phi_2 \> = 0, \quad \< \Phi_1| T_2 |\Phi_2 \> = 0
\ee
where $| \Phi_1\>$ and $| \Phi_2\>$ are physical states. This equivalently boils down to the statement
\be{constr}
L_n | \Phi\> = M_n | \Phi\> = 0, \quad \forall n>0
\ee
where again $\Phi$ is a state in the physical Hilbert space.

We now need to identify the physical Hilbert space. To this end, let us look at the following situation. Suppose we are in a well defined tensile string theory and we are tuning the tension of this theory to zero from a non-zero value. The physical Hilbert space is built on the vacuum state of the tensile string theory which we would call $|0\>_\a$. This is defined by: 
\be{}
 \a_{n}^{\mu}|0\>_\a = 0 = \ta_{n}^{\mu}|0\>_\a \quad \forall n>0.
\ee
Up on quantisation, the tensile theory has states built out of the tensile creation operators acting on this vacuum. The states are of the following form at a general level $(N, M)$:
\be{ph}
| \Phi\> = \a_{-n_1}^{\mu_1}  \a_{-n_2}^{\mu_2} \ldots  \a_{-n_N}^{\mu_N} \ta_{-m_1}^{\nu_1}  \ta_{-m_2}^{\nu_2}  \ldots \ta_{-m_M}^{\nu_M} |0\>_\a 
\ee

We now use the two constraints \refb{constr} on the states that arise from the tensile theory. 
The $L_0$ constraint imposes left-right level matching $N=M$ in \refb{ph}. The $M_0$ constraint gives us the mass of the physical states. 
\be{}
M_0 | \Phi\> =\sum_{m} B_{- m}\cdot B_{m}| \Phi\>=0.
\ee 
The momentum is $P_\mu=\frac{1}{2\pi c'}\dot{X}_\mu$. However the total momentum of the string is:
\be{} p_\mu=\int^{2\pi}_0 P_\mu\ d\sigma=\sqrt{\frac{2}{c'}}B_{0 \ \mu} \ee
Thus we get, 
\be{}  m^2 | \Phi\> =-p^\mu p_\mu | \Phi\> =-\frac{2}{c'}B_0\cdot B_0 | \Phi\> =\frac{2}{c'}\left(\sum_{m\neq 0}B_{-m}\cdot B_{m}\right) | \Phi\> \ee
Here in the last line we have used the $M_0 | \Phi\>=0$ constraint to re-write the sum. The mass of the state $\Phi$ is then given by the following:
\bea{}
m^2 | \Phi\> &=& \frac{2}{c'}\left(\sum_{m\neq 0}B_{-m}\cdot B_{m}\right) \a_{-n_1}^{\mu_1} \ldots  \a_{-n_N}^{\mu_N} \ta_{-m_1}^{\nu_1}\ldots \ta_{-m_N}^{\nu_N} |0\> \\
 &=& \lim_{\e \to 0} \ \frac{2}{c'}\left(\sum_{m\neq 0} \e \eta_{\mu \nu} (\a^\mu_{-m} + \ta^\mu_{m}) (\a^\nu_m + \ta^\nu_{-m}) \right) \a_{-n_1}^{\mu_1} \ldots  \a_{-n_N}^{\mu_N} \ta_{-m_1}^{\nu_1}\ldots \ta_{-m_N}^{\nu_N} |0\> \nonumber
\eea 
This leads to 
\be{} 
m^2 | \Phi\> = 0. 
\ee
We see that we recover (trivially) the expected result that the tensionless sector of a usual tensile string theory has a massless spectrum. We also see that we would be able to generate fields of arbitrary spin which are massless. So in this sense, we have shown that the tensionless limit of string theory generates a theory of massless higher spins. But none of this is very profound in terms of the limit. There is a factor of $\e$ which sits in front of the mass-spectrum and makes all masses vanish. 

\section{The tensionless vacuum}\label{S5}
In this section, we will uncover the most interesting findings of this paper. We will consider tensionless strings as fundamental objects without recourse to any limit and see that we are naturally led to degrees of freedom which are very different from the ones in usual tensile string theory. We shall then look at the mapping between the tensionless theory and the tensile theory in terms of a novel type of Bogoliubov transformation on the string worldsheet and give a description of Rindler-like physics on the worldsheet. We also relate tensionless closed strings to open strings by looking at the truncation of the symmetry algebra on the worldsheet. 

\subsection{States in the tensionless theory}

To this end, we first note that the oscillator construction of the previous sections lead to a form of the mode algebra \refb{AB} which was somewhat unfamiliar when one is concerned about the construction of states on a Hilbert space. To put \refb{AB} into a more familiar looking form, we make the following redefinitions:
\be{C-C}
C^{\mu}_n = \frac{1}{2}({A}^{\mu}_n+B^{\mu}_{n}) \quad \C^{\mu}_n =\frac{1}{2}(-{A}^{\mu}_{-n}+B^{\mu}_{-n}) 
\ee 
or equivalently,
\be{} 
A^{\mu}_n ={C}^{\mu}_n-\C^{\mu}_{-n} \quad B^{\mu}_n ={C}^{\mu}_{n}+\C^{\mu}_{-n}. 
\ee
This means that the Poisson brackets take the usual familiar form:
\be{}
\{C^{\mu}_n, C^{\nu}_m \} = -i n \delta_{n+m, 0} \ \eta^{\mu \nu}, \quad  \{ \C^{\mu}_n,  \C^{\nu}_m \} = -i n \delta_{n+m, 0} \ \eta^{\mu \nu}, \quad \{C^{\mu}_n, \C^{\nu}_m \} = 0.
\ee
The very important point to note here is that the vacuum of the tensionless theory is fundamentally different from that of the tensile theory. The natural tensionless vacuum $(|0\>_C)$ is defined in the tensionless theory with respect to the $C, \C$ oscillators. This is given by
\be{}
C^{\mu}_n |0\>_C = 0 = \C^{\mu}_n |0\>_C \quad \forall n>0
\ee
We shall make the relation between the two vacua $|0\>_C$ and $|0\>_\a$ precise in the next section. For now it is sufficient to understand that the $C$ oscillators are a combination of creation and annihilation operators in the $\a$ language. Hence the two vacua are definitely not the same. 

Let us now describe the states that appear in the fundamentally tensionless sector. These would be states that are built out of the $|0\>_C$ with creation operators. A general state is given by 
\be{psi}
| \Psi\> = C_{-n_1}^{\mu_1}  C_{-n_2}^{\mu_2} \ldots  C_{-n_N}^{\mu_N} \C_{-m_1}^{\nu_1}  \C_{-m_2}^{\nu_2}  \ldots \C_{-m_M}^{\nu_M} |0\>_C
\ee
One can easily rework the computations of the previous section in terms of the $C, \C$ oscillators. The expressions for the classical constraints become
\bea{} && L_n=\sum_{m} (C_{-m}\cdot C_{m+n}-\C_{-m}\cdot \C_{m-n})=0  \\ 
&& M_n=\sum_{m} (C_{-m}\cdot C_{m+n}+\C_{-m}\cdot \C_{m-n}+2C_{-m}\cdot \C_{-m-n})=0 \eea
The constraints on the physical states are as usual:
\be{}
L_n |\Psi \> = 0, \quad M_n |\Psi\>=0 \quad \forall n>0.
\ee
The $L_0$ constraint now gives us 
\be{} L_0 |\Psi \> =\sum_{m} (C_{-m}\cdot C_{m}-\C_{-m}\cdot \C_{m}) |\Psi \> =(N_C - N_{\C}) |\Psi \> = 0 \ee
which implies that the physical tensionless states should have equal number of $C$ and $\C$ oscillators. 

The interesting constraint is the $M_0$ constraint. Like before, we can compute the momentum as $P_\mu=\frac{1}{2\pi\alpha'}\dot{X}_\mu$. The total momentum of the string is:
\be{} p_\mu=\int^{2\pi}_0 P_\mu\ d\sigma=\sqrt{\frac{2}{c'}}(C_{0 \ \mu}+\C_{0 \ \mu}) \ee
Thus, 
\be{tm} m^2 |\Psi \>  =-p^\mu p_\mu |\Psi \> =-\frac{2}{c'}(C^\mu_0+C^\mu_0)^2  |\Psi \> =\frac{2}{c'}\sum_{m\neq 0}(C_{- m}\cdot C_{m}+\C_{- m}\cdot \C_{m}+2C_{- m}\cdot \C_{-m}) |\Psi \> 
\ee
The last term in the above equation is problematic. This means that these states don't have well defined mass. It is abundantly clear from \refb{tm} that the states that we have constructed out of the tensionless vacuum are very unlike the states in the usual tensile theory Hilbert space.

\subsection{Tensile vacuum to tensionless vacuum: Bogoliubov transformations}

We have in previous sections seen how to map between the oscillators $\a, \ta$ and $A, B$. The exact mapping is given by \refb{ABmap}. Combining this with \refb{C-C}, we find that the tensionless oscillators $C, \C$ are related to $\a, \ta$ by 
\bea{bogo}
C^{\mu}_n(\e) &=&\beta_+ {\alpha}^{\mu}_n+\beta_-\tilde{\alpha}^{\mu}_{- n} \\
\C^{\mu}_n(\e) &=&\beta_- {\alpha}^{\mu}_{-n}+\beta_+\tilde{\alpha}^{\mu}_{ n} \nonumber
\eea
where 
\be{}
\beta_\pm=\frac{1}{2}\left(\sqrt{\e}\pm\frac{1}{\sqrt{\e}}\right).
\ee 
Since the canonical commutation relations are preserved in going from the $\a$ basis to the $C$ basis, the above equations define a Bogoliubov transformation {\em{on the string worldsheet}}. 

It is easy to see that $\e=1$ just lands one on the usual tensile string theory $(C^{\mu}_n= \a^\mu_n, \ \C^{\mu}_n= \ta^\mu_n)$, while as usual, the $\e \to 0$ limit is the tensionless limit. The parameter $\e$ runs from 1 to 0. As one dials $\e$ from 1 to 0, the effective string tension becomes smaller and smaller and in the exact limit one lands on the tensionless string. We will now be interested in the whole range of $\e$ and not only in the strict limit. 

We can redefine the coefficients in a form more familiar to applications of Bogoliubov transformations \cite{Blasone}:
\be{}
\beta_+ = \cosh \theta, ~~ \beta_-= \sinh \theta,
\ee
With this new defination of coefficients the Bogoliubov transformation can be recasted as,
\bea{CC}
C^{\mu}_n = e^{-i G} \alpha_{n} e^{iG} =\cosh\theta \ {\alpha}^{\mu}_n -\sinh\theta \ \tilde{\alpha}^{\mu}_{- n}   \\
\tilde{C}^{\mu}_n= e^{-i G} \tilde{\alpha}_{n} e^{iG}=-\sinh\theta \ {\alpha}^{\mu}_{-n}+\cosh\theta \ \tilde{\alpha}^{\mu}_{ n} \nonumber
\eea
The generator of the above transformation can be written as,
\be{G}
G = i \sum_{n=1}^{\infty} \theta \Big[ \alpha_{-n}.\tilde{\alpha}_{-n} - \alpha_n .\tilde{\alpha}_n\Big],
\ee
where all the spacetime indices of the oscillators are contracted. Now we are in a position to write down the mapping between the two vacua $|0\>_\a$ and $|0\>_C$. This is given by the following
 \begin{eqnarray}
 |0\>_C &=&  \exp[i G] |0\>_\a \nonumber \\
 &=&  \exp \Big[- \theta \sum_{n=1}^{\infty} [ \alpha_{-n}.\tilde{\alpha}_{-n} - \alpha_n .\tilde{\alpha}_n]\Big] |0\>_\a \nonumber \\
 &=& {\left(\frac{1}{\cosh\theta}\right)}^{1+1+1+1+\hdots\infty}~~~\prod_{n=1}^{\infty}  \exp[\tanh\theta \alpha_{-n}\tilde{\alpha}_{-n}] |0\>_\a 
 \end{eqnarray}
 We now regulate our answer by replacing the diverging sum: $1+1+1+\hdots\infty=\zeta(0)=-\frac{1}{2}$ in the usual way to get:
 \be{vac}
{\boxed{ |0\>_C= \sqrt{\cosh\theta} \prod_{n=1}^{\infty}  \exp[\tanh\theta \alpha_{-n}\tilde{\alpha}_{-n}]  |0\>_\a}}
 \ee
In terms of the original vacuum $|0\>_\a$ and its oscillators $\a$, the new vacuum $|0\>_C$ is a {\em{coherent state}}, or more precisely a {\em{squeezed state}}. This is hence a highly energised state with respect to the old vacuum. 

As is very well known, when one looks at accelerated observers in Minkowski spacetimes, these Rindler observers detect particles in their vacuum state. In close analogy to Rindler physics, the new string theory vacuum state would be one which is also bubbling with particles. It is easy to observe this by computing the expectation value of the left and right handed number operator for any given mode:
 \be{part}
 {}_C\< 0| n^L_k |0\>_C = {}_C\< 0|a_{-k} a_{k}|0\>_C = \sinh^2\theta  \quad \text{where} \quad a_k = \sqrt{k} \ \a_k 
 \ee
Similarly ,
 \be{}
{}_C\< 0| n^R_k|0\>_C = {}_C\< 0|\tilde{a}_{-k} \tilde{a}_{k}|0\>_C = \sinh^2\theta   
\ee
where again the operators $\tilde{a}$ have been defined as above \refb{part}. Similar to Rindler physics, the observer who is in the $|0\>_C$ vacuum would always observe a thermal spectrum and effectively be at a finite temperature. We will provide some more arguments in this direction in the next section. It is good to see that when you look at the case of the tensile theory $\e=1$, then $\sinh \theta =0$ and there are no particles in the ground state $|0\>_\a$. 

Before moving on, it is of interest to take stock of the situation at hand. We have argued in previous sections that when one looks at the tensionless string theory, there is the emergence of a new vacuum state. This vacuum is the one which is natural in the fundamentally tensionless theory. The states in the fundamentally tensionless theory are excitations created by $C$ creation operators on this new ground state $|0\>_C$. We have seen that these states are very different from usual tensile states; they are not mass eigenstates. So we have seen the emergence of new degrees of freedom in the tensionless limit. Above we have looked at a situation where one dials the string tension from the usual value to zero. We have seen in this situation, we can find a mapping between the two Hilbert spaces in terms of Bogoliubov transformations, especially we can view the new vacuum as a coherent state in terms of the usual tensile theory. This gives a manifestation of Rindler-like physics on the string worldsheet.  

\subsection{Relating tensionless closed strings and open strings}

In this subsection, we wish to report on another very interesting feature which also emerges directly from the construction described earlier. As we have seen earlier, the classical constraints can be rewritten as operator equations on the Hilbert space of the tensionless string: 
\be{}
L_n |\Psi \> = M_n |\Psi \> = 0 \quad \forall n>0
\ee
For the zero modes, we can have additionally some normal ordering constant
\be{}
L_0 |\Psi \> = a_L |\Psi \>, \quad M_0 |\Psi \> = a_M |\Psi \> 
\ee
This would mean that the physical states are primary operators of the 2d GCFT with $(h_L, h_M) = (a_L, a_M)$. 

If we are restricting ourselves to tensionless string theories arising out of tensile theories, in which case $|\Psi\>$ is a state in the Hilbert space of the tensile theory as in our discussions around \refb{ph}. In this case, due to the limit, we would be lead to a sector where $h_M=0$. It was further argued in \cite{Bagchi:2013bga} that if this tensionless theory arises from a theory with a finite number of worldsheet fields, $c_M = \e (c + \bar{c}) = 0$. 

In the fundamentally tensionless case, there seems to be the possibility of having non-zero values for all of $\{ h_L, h_M, c_L, c_M \}$. For the moment, let us focus on tensionless theories which arise as a limit from well defined tensile theories.  The physical operators which are used to build physical states in the theory have $h_M=0$ as argued above and this is also a theory which has vanishing $c_M$. It can be shown through an analysis of null vectors that a 2d GCFT with $h_M=0, c_M=0$ is equivalent to a chiral half of a CFT with the symmetries of a single copy of the Virasoro algebra, or in other words there is a consistent truncation of the representations of the 2d GCA to the representations of a single copy of the Virasoro algebra in this sector \cite{Bagchi:2009pe} and is called the chiral truncation of the GCA \cite{Bagchi:2012yk}. The analysis of this phenomenon using null vectors of the GCA is presented in the appendix \ref{ApA} for completeness. 

This seems to have very interesting physical significance. A theory of strings with one copy of the Virasoro algebra as its residual symmetry is obviously reminiscent of open strings. We thus see that closed tensionless strings rather naturally behave as open strings. This strange behaviour of the tensionless strings has long been expected (see e.g. \cite{Francia:2007qt, Sagnotti:2011qp}) but, to the best of our knowledge, never been explicitly demonstrated. 

The chiral truncation of the 2d GCA has been previously used to formulate an exact correspondence between a theory of asymptotically gravity in 3 dimensions and a chiral half of a CFT in 2 dimensions, providing the first concrete example of holography in flat spacetimes \cite{Bagchi:2012yk}. Here we find that similar arguments on the worldsheet lead us from a theory of closed tensionless strings to open strings. 

Let us comment on the fundamental tensionless string theory before closing this section. One could argue against setting $c_M=0$ for the fundamentally tensionless string and say that this only arises in a certain sub-sector of the theory. We would like to point out that if one imposes unitarity on the highest weight representations, following \cite{Bagchi:2009pe, Bagchi:2012yk, Grumiller:2014lna} one naturally lands in the sub sector with $h_M =0, c_M=0$. So, demanding unitarity of the underlying symmetry structure leads us to conclude that closed strings behave like open strings in the tensionless sector. 

\newpage

\section{Applications: Hagedorn physics and more}\label{S6}
Having discussed so far a rather exotic limit of string theory, from the point of view of worldsheet symmetries, it is now important to put things in perspective and discuss where the construction that we have rather meticulously put together so far may be useful. To this end we discuss first some aspects of strings at very high temperatures and then some other possibly interesting situations. We should stress that the comments in this section are essentially preliminary remarks and all of the points discussed would need to be backed up with more substantial calculations, which we hope to present in upcoming work. 

\subsection{Strings near Hagedorn temperature}

As discussed in the introduction, the extreme high energy limit of string theory is of interest due to the possible existence of phase transitions between the normal and Hagedorn phase. In string theory, one encounters an exponential growth in the single-string density of states as a function of mass. The number of states at the $n$th level grows approximately like exp$(4 \pi n^{1/2})$ (see e.g. \cite{Green:1987sp, Polchinski:1998rq}). This rapid growth makes the canonical partition function of a free string gas 
\be{}
Z_{\text{string}}= \mbox{Tr} \ e^{- \beta H}
\ee
well-defined only at sufficiently low temperatures. The partition function diverges at temperatures above
\be{}
T_H = \frac{1}{4 \pi {\sqrt{\alpha'}}}
\ee
where $T_H$ is the Hagedorn temperature. This has lead to the suggestion that the Hagedorn temperature defines an absolute limiting temperature in physics or that the phenomenon signals a transition to a different and unknown phase governed by fundamental degrees of freedom which are very different from usual strings. 

Following e.g. \cite{Green:1987sp}, we can compute the high energy limit of the single string density of states for free closed strings and this is found to be 
\be{}
\omega(\e) \sim V e^\frac{\e}{T_H} \e^{- d/2 -1}
\ee
where V is the volume of the system and $d$ is the number of non-compact space directions. The energy distribution function $D(\e,E)$ in the micro-canonical ensemble gives the average number of strings characterised by the range of energies between $\e$ and $\e + \delta \e$ constrained to the total energy $E$. This can be calculated from the above expression for the density of states and it can be shown that at high energy density that the thermodynamically favourable configuration is a single long string carrying the bulk share of the available energy \cite{Bowick:1989us, Giddings:1989xe, Lowe:1994nm}.

The above picture is made more complicated when one includes interactions. In the thermodynamic limit, the above mentioned long string will traverse the entire system many times over and intersect itself a large number of times. String interactions would then affect the equilibrium configuration and this was investigated in \cite{Lowe:1994nm} where it was found that the long string phase would still exist which would be dominated by a large number of long strings which may split and join. 

The tensionless limit of string theory is intimately related to the physics at the Hagedorn scale as the effective string tension at $T_H$ goes to zero \cite{Pisarski:1982cn, Olesen:1985ej}. The effective string tension $\T_{\text{eff}}$ is given by 
\be{}
\T_{\text{eff}} = \T \sqrt{1 - \frac{T^2}{T_H^2}}
\ee
So, here $\e=\sqrt{1 - \frac{T^2}{T_H^2}}$ , the parameter that we have been using all along to get to the tensionless string from the tensile string. The analysis of the previous sections thus find a very natural home in the theory of strings near the Hagedorn temperature. We have found that there is the emergence of degrees of freedom which are very different from the usual tensile theory. We have also discovered a new vacuum state $|0\>_C$ that differs significantly from the tensile vacuum $|0\>_\a$. As we have seen, the new vacuum state is a coherent state in terms of the tensile theory and is hence a highly energised state. We propose that this new vacuum state $|0\>_C$ {\em{is the worldsheet manifestation of the emergent long string at the Hagedorn temperature}}. 

The excitations around this new vacuum state are described by the creation operators $C_i, \C_i$ acting on $|0\>_C$. We saw in the earlier sections that these excitations are very unlike the usual string states, e.g. they don't have a proper definition of mass. From the point of view of the underlying algebra, this arises from the fact that when one looks at the residual symmetry of the tensionless string, in the highest weight representation, the operator $M_0$ has a Jordon-block structure. There are numerous claims in the literature that the fundamental degrees of freedom drastically change when one approaches the Hagedorn temperature. We see from our analysis that from the worldsheet there is the emergence of new degrees of freedom in terms of the excitations of the new vacuum state and we interpret these as the above mentioned mysterious degrees of freedom near the Hagedorn temperature which emerge when one excites the long string by pumping in further energy into the system. 

It is often claimed that the Hagedorn temperature is not a limiting temperature but an indication of a phase transition. Here one appeals to the QCD story where hadrons can be thought of as string-like electric flux tubes and hence low energy QCD can also be described in terms of strings. When QCD is heated up, one similarly finds a limiting temperature because the degeneracy of states of these flux tubes increases in the same fashion described above for the fundamental string. This ``limiting" temperature is however simply an indication of the deconfining phase transition where the QCD strings disintegrate into a quark gluon plasma. In string theory, the reason for the divergences at the Hagedorn temperature is that a particular string mode becomes massless, indicating a second order phase transition \cite{Sathiapalan:1986db}. It was argued in \cite{Atick:1988si} that before the second order transition can occur, due to the coupling to dilatons an instability arises leading to a first order transition. In flat space in the thermodynamic limit of infinite volume, it is possible for the string to exist in the metastable phase and hence be sensitive to the Hagedorn temperature even though there is a first order transition. 

We don't have a lot to say about the interacting picture in our present work. But one of the claims usually presented in the above arguments is that the structure of the string worldsheet breaks down near the Hagedorn temperature. What we would like to advocate through our analysis is that (when one looks at the non-interacting case) the description of the string in terms of the worldsheet does not break down, but changes in nature. We have seen that in the tensionless limit, we don't have a notion of a worldsheet metric any more. This is degenerate. The usual string worldsheet is a Riemannian manifold with a well-defined metric. The tensionless worldsheet, which is what the worldsheet becomes near the Hagedorn temperature, is now replaced by a Carrollian manifold \cite{Duval:2014uoa, Duval:2014lpa}. Spacetimes where the speed of light has been taken to zero are examples of Carrollian manifolds. These are closely related to (or more accurately dual to) Newtonian-Cartan manifolds. In two dimensions, as is the case we are dealing with, Carrollian and Newtonian-Cartan manifolds lead to isomorphic symmetry structures. 

It has long been suspected that the structure of spacetime itself would change when one considers strings near the Hagedorn temperature. Above we have argued that the structure of the worldsheet manifold changes quite dramatically from Riemannian to Newton-Cartan. This would induce a change in spacetime as well. In \cite{Duval:2014lpa}, the authors claim that the tensionless strings move in what are known as Bargmann spacetimes. It would be of great interest to investigate this in further detail and understand what sort of modification the original Minkowski spacetime undergoes when one looks at the tensionless string. We hope to be able to understand the deformations of the spacetime geometry by strings at high temperatures by pushing this formalism further. 

\subsection{Entanglement on the worldsheet and thermal vacuum}

Before moving on to other potential applications, we want to make another possibly interesting suggestion. This is to do with the notion of temperature on the string worldsheet. We have said that the usual string vacuum and the string states build above it with the $\a, \ta$ constitute what is the low temperature phase. The other vacuum $|0\>_C$ and its excitations build up the high temperature phase.  One way to understand the thermal phase is to invoke the notion of the Thermo Field Double \cite{Takahasi}. This formalism replaces thermal averages by vacuum expectations in a suitable Fock space:
\be{Atfd}
\< A \> = Z^{-1} (\beta) \ \text{Tr} (e^{-\beta H} \ A) = \< 0(\b) |  A  | 0(\b) \>
\ee
where $A$ is a generic observable. The thermal vacuum $ | 0(\b) \>$ is a state that satisfies
\be{}
\< 0(\b) |  A  | 0(\b) \> = Z^{-1} (\beta) \sum_n \<n| A |n\> e^{-\b E_n}
\ee
Here $|n\>$ are energy eigenstates of the underlying Hamiltonian $H |n\> = E_n |n\>$. These are also orthonormal $\<n|m\> = \delta_{n,m}$. Expanding in terms of $|n\>$, the thermal vacuum is given by
\be{}
| 0(\b) \> = \sum_n f_n( \b) |n\> \quad \text{where} \quad f^*_n( \b) f_m( \b) = Z^{-1} (\beta) e^{-\b E_n} \delta_{n,m}
\ee
$f_n( \b)$ are thus clearly not numbers, but the relation above is like the orthonormality of vectors. If we consider a larger Hilbert space, obtained by doubling of the degrees of freedom,  $|0(\b)\>$ can be thought of as living in this space. The construction has necessitated the introduction of a fictitious dynamical system identical to the one under consideration: 
\be{} \tilde{H} |\tilde{n}\> = E_n |\tilde{n} \>	 \quad \< \tilde{m}|\tilde{n} \> = \delta_{n,m}
\ee
The energy is postulated to be the same of the one of the usual states. The thermal vacuum is thus given by 
\be{}
|0(\b) \> = Z^{-\frac{1}{2}}(\b) \ \sum_n e^{- \frac{\b}{2} E_n} \ |n\> \otimes  |\tilde{n} \>
\ee
This thermo-field double formalism has been used by Maldacena to construct a CFT dual of an eternal AdS black hole \cite{Maldacena:2001kr}. This has been popular of late in the extended debate on firewalls and calculations of entanglement entropy. 

In \refb{Atfd}, $A$ is an operator on the usual Hilbert space. The first step of \refb{Atfd} does not contain any operators on the tilde Hilbert space and hence we can sum over all states of the tilde copy. Doing this leads to a thermal density matrix for the original copy of the field theory. The thermal density matrix arises out of entanglement and the entropy is the entanglement entropy. 

In close analogy with the above, we propose that our thermal vacuum $|0\>_C$ arises out of entanglement between the right movers and the left movers of the 2D worldsheet residual conformal symmetry. The underlying assumption is that the specific operators that we would be dealing with are holomorphically factorisable. In order to make this connection possible, let us revisit some of the earlier statements. In the tensile string, we have two sets of commuting creation and annihilation operators:
\be{}
[\a_n, \a_m] = n \delta_{n+m}, \quad [\ta_n, \ta_m] = n \delta_{n+m}
\ee
where we have suppressed the spacetime indices. The Hamiltonian of the left-moving (`un'-tilde) sector is given by $H_L = \sum \a_{-n}\cdot \a_n$ and similarly for the right moving (tilde) sector $H_R = \sum \ta_{-n}\cdot \ta_n$. The thermal operators $C_n, \C_n$ are defined by \refb{CC}. In the present context, $\theta = \theta(\b)$ which is a function to be determined. And the generator of the transformation $G$ is given by \refb{G}. 

To identify the new vacuum state $|0\>_C$ with the thermal vacuum, we need to find an explicit relation between $\theta$ and $\b$. To this end, we look at the usual thermo field dynamics procedure of minimising the potential 
\be{}
F= E - \frac{1}{\b} S
\ee
with respect to the parameters of the Bogoliubov transformation. Hence F is a free energy like potential. In this formalism, the thermal energy is given by computing the matrix elements of the $T = 0$ Hamiltonian in the thermal vacuum. The entropy $S$ is computed similarly by using the entropy operator $\hat{S}$ defined for the left-movers of the closed string by:
\be{}
\hat{S} = - \sum_n \left[a_k^{\dagger} a_k \log\sinh^2{\theta} -  a_k a_k^{\dagger} \log\cosh^2{\theta} \right]
\ee
Here 
\be{}
\a_n = \frac{a_n}{\sqrt{n}}, \quad \a_{-n} =  \frac{a^{\dagger}_n}{\sqrt{n}}, \quad n>0. 
\ee
The action of $\hat{S}$ can be interpreted as tracing over the right-moving degrees of freedom and hence the entropy S is the entanglement entropy on the worldsheet of $|0(\theta)\>$ after tracing out the right-sector. Looking again at the left sector, the Hamiltonian is 
\be{}
H = \sum_n \omega_n a^\dagger a, \quad \text{where} \quad \omega= n+ \frac{1}{2}
\ee
We wish to evaluate
\be{}
\Omega = \< 0 (\theta) | F | \ 0 (\theta) \>  = \< 0 (\theta) | \frac{1}{\b}\left(a_k^{\dagger} a_k \log\sinh^2{\theta} -  a_k a_k^{\dagger} \log\cosh^2{\theta}\right) + \omega_n a^\dagger a | \ 0 (\theta) \>
\ee
Minimising this using \refb{part} we get, 
\be{}
n^L_k = \frac{1}{e^{\b (n+\frac{1}{2})} -1}
\ee
We thus see that we indeed have a thermal Bose distribution and can conclude that our new vacuum $|0\>_C$ is indeed the thermal vacuum $|0 (\b)\>$ and is generated in this doubled formalism by means of the Bogoliubov transformations which results in the entanglement of the left and right Virasoros on the string theory worldsheet.   

Our proposed formalism of entanglement between the left and right sectors of the 2D worldsheet CFT is similar to some recent discussions of left and right entanglement in spacetime 2D CFTs \cite{PandoZayas:2014wsa, Das:2015oha}. It is interesting to note that the thermal vacuum that we have found here are similar to the boundary states that are considered in these references. The fact that we have also found that closed tensionless strings behave like open strings makes this connection even more tantalizing. We would like to investigate these issues in future work. 

\subsection{Other applications}

The tensionless limit of string theory is also of interest when one considers strings in strong gravitational fields \cite{DeVega:1992tm}. Strings behave like classical tensionless objects in the vicinity of spacetime singularities. Here again, there is the increase in length of the string as one approaches the singularity and once again, the emergent vacuum state $|0\>_C$ would be important in understanding the dynamics. 

We also expect our analysis to be of use for strings which approach black hole horizons and get stretched to long strings \cite{Halyo:1996vi}. In this context, we also hope that our construction would be of interest to the recent attempt at solving some problems of quantum black holes using long strings and tensionless branes advocated in \cite{Martinec:2015lka}. 

Our constructions of tensionless strings should apply to situations outside string theory and in particular in the theory of QCD. There has been recent work linking the deconfinement transition of large-$N$ Yang-Mills theory to the condensation of very long chromo-electric flux strings \cite{Hanada:2014noa}. This has been argued to be analogous to the formation of a black hole in string theory. Although in this picture, interactions of the intersecting QCD strings play an important role, it is very plausible that the worldsheet picture could provide interesting hints to the same physics.

\section{Conclusions and future directions}\label{S7}

\subsection{Summary of results}
In this paper, we have looked closely at the theory of tensionless strings from the point of view of the symmetries of the string worldsheet. We have treated the tensionless string both as a limit from the usual tensile theory and as a fundamental object. We have found that results in these two different ways of approaching the tensionless string agree in the classical analysis. The expressions for the equations of motion, mode expansions, constraints are all compatible with the algebraic structures arising out of the underlying residual gauge symmetry governed by the Galilean Conformal Algebra. It was particularly gratifying to see the emergence of the non-obvious form of the EM tensor of the 2D GCFT from the constraints. We also recovered all of this analysis by carefully looking at the ultra-relativistic limit on the string worldsheet. 

Significant differences in the limiting and the fundamental approach become apparent when studying the quantum tensionless strings. Here we found that the fundamentally tensionless theory lead us to a vacuum which is clearly unlike the vacuum of the tensile theory. We then distinguished between two distinct approaches, the limiting one and the fundamental one. In the case where one deals with the tensionless limit of a particular well-defined string theory, the physical states are given by the states of the original string theory. Applying the constraints generated by the tensionless limit, we found, rather trivially, that the masses of all these states became zero and this is in our analysis the emergence of the infinite tower of massless particles of arbitrarily high spin. 

The natural vacuum in the tensionless theory, is however very different. By looking at worldsheet Bogoliubov transformations, we found that this new vacuum could be looked at as a highly energised coherent state in terms of the tensile vacuum and the tensile oscillators. We believe that is the most striking result of our paper. As a realisation of the calculations of our work, we looked at the theory of strings near the Hagedorn temperature. As one pumps energy into a gas of strings, near the Hagedorn temperature, it becomes thermodynamically favourable to form a single long string and this is the dominant phase. We proposed that the tensionless vacuum that we constructed as a coherent state out of usual tensile operators was the worldsheet manifestation of the long-string phase near the Hagedorn temperature. The excitations around this vacuum generate the often discussed new degrees of freedom in the Hagedorn phase which are very unlike usual string states. 

Some of the other interesting results of the paper are the following. We saw that tensionless closed strings behave very much like open strings as there is a truncation of the GCA to a single copy of the Virasoro algebra for the vanishing of a particular central term ($c_M=0$). This is direct evidence of a property which has been hinted at before but never demonstrated. 

In the discussion relating Hagedorn physics and tensionless strings, we also proposed a novel way of generating the thermal vacuum (or the long string) in terms of entanglement between the two copies of the Virasoro algebra of the 2d CFT on the worldsheet. 

\subsection{Connections to flat holography}

We have seen the physics of tensionless strings is governed by the underlying structure of the residual gauge symmetry, the 2d GCA. The most natural limit is the worldsheet ultra-relativistic 
limit. As stated before, the construction of holography for flat spacetimes is linked to the same limit. To elaborate a bit further, the essential idea that has been pursued in the field of flat holography builds on the fact that flat space is obtained as an infinite radius limit of AdS spacetimes. It is thus natural to look at flat holography as a limit of AdS/CFT. The initial progress on flat holography has been principally focussed on the 3D bulk and 2D boundary theory. On the side of the dual field theory, the infinite radius bulk limit translates to an ultrarelativistic limit which contracts the 2d CFT to a 2d GCFT, giving rise to a Flat/GCFT correspondence. This was originally dubbed the BMS/GCA correspondence in \cite{Bagchi:2010eg} following the observation of the isomorphism between the 2d GCA and the asympototic symmetries of a 3D Minkowski spacetime, the BMS algebra. 

Our analysis in the present paper throws up interesting suggestions about flat holography as well. The construction of the tensionless vacuum is inherently linked to the discussions of GCFT and hence this could be a hint to understanding how fields and operators in AdS map over to the corresponding ones in flat space. For holographic discussions, the highest weight representations of the corresponding asymptotic symmetry algebras have always played a crucial role. So one needs a mapping between the highest weight representations of the Virasoro algebra and that of the 2d GCA. But the ultra-relativistic limit \refb{ur-lim} mixes the creation and annihilation operators in the mapping and hence under the UR map, the Virasoro highest weights don't go over to the GCFT highest weights, providing a conceptual hurdle in mapping some aspects of physics in AdS to that in flat space. 

But the discovery in the present paper of the Bogoliubov transformations between the oscillators of the tensile and tensionless theory should pave the way to an answer to the above problem. In particular, it is plausible that the Minkowski ground state is a coherent state in terms of the AdS vacuum.

\subsection{The road ahead}

As should be obvious from the numerous discussions above, there are a large number of avenues to pursue in future work. Let us systematically list a few of them. 

\paragraph{What happens to spacetime?} We have looked at the tensionless limit of string theory by focussing on the worldsheet aspects of the theory. An obvious question is what effect does this have on the spacetime. We had started out with closed bosonic strings propagating on 26 dimensional Minkowski space. The limit on the worldsheet would naturally induce a deformation of the spacetime. As stated in Sec~5, \cite{Duval:2014lpa} indicates that tensionless strings propagate on Bargmann spacetimes. It would be of great interest to see this more explicitly and understand the deformation of the initial Minkowski space. We should be able to arrive at similar considerations by looking at the quantum version of Galilean Conformal invariance on the tensionless worldsheet, which should be analogous to the vanishing of the (relativistic) beta functions required for maintaining (relativistic) conformal invariance on the tensile string worldsheet. This would be our immediate goal. 

\paragraph{Open strings and D-branes:} We have looked at closed bosonic strings in our construction and not ventured into open string theory. As was discussed in \cite{Bagchi:2013bga}, from the point of view of the symmetry algebra, it is far from obvious how to contract a single Virasoro algebra, as is the case of the open string residual symmetry, to get something non-trivial. One needs to develop the theory of Boundary Galilean Conformal Field Theories to get a handle on this. It would be easier to first do the intrinsic analysis of the open string, arrive at the results and then try and interpret what the precise limit on the open worldsheet implies. This is work in progress. It would also be of importance to understand how the tensionless limit works for branes. Here we would not have the luxury of the systematics of GCFT to fall back to. 

\paragraph{Superstrings and beyond:} It is of course natural to look at closed superstring theories as an immediate generalisation. We expect that the construction would be straight-forward and the residual symmetry would also be a contraction of the two copies of the super-Virasoro algebra to the Super-GCA. The construction of the map between the tensionless and tensile operators would now also involve fermionic generators and the expressions of the coherent state would change appropriately. 

\paragraph{Hagedorn Physics:} We have only skimmed the surface and made a couple of suggestions about how the construction of tensionless strings would be of importance to strings near the Hagedorn temperature. It is vitally important to back this up with more concrete calculations. For example, we wish to perform partition function calculations with the new vacuum and the excitations above it to show how the Hagedorn divergence is tamed when one looks at these emergent degrees of freedom. It would also be of great importance to back up our picture of the thermal vacuum state being constructed out of entanglement between left and right movers on the string worldsheet. 

\bigskip

To conclude, we would like to say that we seem to have unearthed some interesting new effects by studying systematically the theory of tensionless strings from the point of view of the worldsheet symmetries. There are a great many adventures that lie ahead and many potential applications of our current results.

\section*{Acknowledgements}
It is a pleasure to thank Tarek Anous, Aritra Banerjee, Rudranil Basu, Diptarka Das, Mirah Gary, Rajesh Gopakumar, Daniel Grumiller, Hong Liu, Sudipta Mukherji, Alfonso Ramallo, Balachandra Sathiapalan and Joan Simon for discussions. The work of AB is supported by the Fulbright Foundation. SC and PP are supported by Erasmus Mundus NAMASTE India-EU Grants. AB thanks the Vienna University of Technology, University of Santiago de Compostela, Harish Chandra Research Institute, Universite de Libre Brussels for hospitality during various stages of this work. SC thanks the Institute of Mathematical Sciences and the Institute of Physics for hospitality.

\appendix

\section{Tour of Galilean conformal field theories} \label{ApA}
We provide a quick review of the basic features of Galilean Conformal Field Theories in this appendix. Much of our understanding of the tensionless limit of string theory would be based on algebraic structures that arise from this symmetry. So to make the paper self-contained, at the cost of repeating ourselves for the experienced reader, we discuss the main points of GCFTs, especially GCFTs in $D=2$. 

\subsection{Galilean conformal symmetries in general dimensions}

GCFTs were first understood as the symmetries of non-relativistic conformal symmetries. The most intriguing feature of GCFTs is that, unlike relativistic CFTs, the underlying symmetry is infinite dimensional in all space-time dimensions. Algebraically, the set of vector fields that generates these symmetries can be denoted by:
\be{gcaD}
L_n = t^{n+1} \p_t + (n-1) t^n x_i \p_i, \quad M^i_n = t^{n+1} \p_i
\ee
where $n$ runs over all integral values. These vector fields follow the algebra
\be{}
[L_n, L_m] = (n-m) L_{n+m}, \quad [L_n, M^i_m] = (n-m) M^i_{n+m}, \quad [M^i_n, M^j_m]= 0. 
\ee
Together with this, there are also the generalised rotation generators
\be{}
J_{ij}^n = t^n \left(x_i \p_j - x_j \p_i\right)
\ee
which lead to current-algebra like commutators with the rest of the algebra \cite{}. We would not be interested in this here since we are looking to focus on two dimensions. 

Before moving to $D=2$, we would like to point out some important points. Firstly, the finite dimensional subgroup $\{L_{0, \pm1}, M^i_{0, \pm1}, J^0_{ij} \}$ of the above Galilean Conformal Algebra arises from the contraction of the relativistic conformal algebra in $D$ dimension. The vector fields in \refb{gcaD} for $n,m=0, \pm1$ correspond to 
\be{}
L_{-1, 0, +1} = H, D, K^0 \qquad M^i_{-1, 0, +1} = P^i, B^i, K^i 
\ee
where $K^0, K^i$ are the contracted versions of the temporal and spatial components of the special conformal transformation and $B^i$ are the Galilean boosts. $H, D$ are the Hamiltonian and the Dilatation operator while $P^i$ are the generators of spatial momenta. The infinite enhancement is conjectured by grouping the generators of the contracted algebra in the specific way mentioned above and by noting that the vector fields close to form the same algebra for all integral values of the modes. 

Secondly, since the infinite dimensional extension of symmetries in the non-relativistic limit is conjectural and seems to be rather counterintuitive, it is essential that we present some examples to back up the claim. These extended symmetries are indeed realised in physical systems, viz. the Naiver-Stokes' equation of non-relativistic hydrodynamics realises all the time dependent boosts ($M_n^i$'s) as symmetries \cite{Bagchi:2009my}. More recently, it has been shown that the equations of motion of Maxwell electrodynamics in the non-relativistic limit exhibits {\em{the entire}} Galilean Conformal Algebra as their symmetry \cite{Bagchi:2014ysa}. Current investigations show that this also extends to Yang-Mills theories and indeed one can hope for a general proof that all CFTs in the non-relativistic limit will exhibit this extended infinite dimensional symmetry.     

\subsection{Galilean conformal symmetries in $D=2$}

Much of our explorations in this paper would involve the details of GCFTs in $D=2$. The generators of the GCA in $D=2$ involves the $L_n$'s and the $M_n$'s (with the vectorial index dropped) and, obviously, there are no $J_{ij}$'s due to the absence of spatial rotation in $D=2$. The quantum version of the GCA is given by:
\bea{gca2d}
&&[L_n, L_m] = (n-m) L_{n+m} + \frac{c_L}{12}(n^3-n) \delta_{n+m,0} \, , \cr
&&[L_n, M_m] = (n-m) M_{n+m} + \frac{c_M}{12}(n^3-n) \delta_{n+m,0} \, , \quad [M_n, M_m]= 0. 
\eea
It should not be a surprise that the infinite dimensional 2D GCA can be obtained as a contraction of two copies of the Virasoro algebra. If we define the relativistic conformal algebra by $\L_n, \bL_n$, the following combination of the Virasoro generators give rise to the 2D GCA in the limit $\e \to 0$:
\be{nr-lim}
%\hspace{-3cm}\mbox{Non-Relativistic Limit:} \quad  
\L_n + \bL_n = L_n, \quad \L_n - \bL_n = \frac{1}{\e} M_n
\ee
The central terms $c_L, c_M$ in the quantum 2d GCA \refb{gca2d} can be linked to the original CFT central terms in this non-relativistic limit:
\be{c-nr}
c+ \bar{c} = c_L, \quad c- \bar{c} = \frac{1}{\e} c_M
\ee
It is instructive to note that by looking at the representations of the Virasoro generators on the complex plane, viz. 
\be{}
\L_n = z^{n+1} \p_z, \quad \bL_n = \z^{n+1} \p_\z
\ee
and taking $z = t + \e x, \, \z = t -\e x$, one can readily obtain the 2d version of the GCA generators \refb{gcaD} in the limit $\e \to 0$. One can also look at the limit on the cylinder $z= e^{i \w}, \, \z= e^{i \bw}$. Contraction of the generators then leads to 
\be{cyl-NR}
L_n = e^{in\t} \left( \p_\t + i n \s \p_\s \right), \quad M_n = e^{in\t} \p_\s
\ee
where we have written $\w, \bw = \t \pm \e \s$ and taken the limit $\e \to 0$. In purely non-relativistic terms, the mapping between the ``plane" co-ordinates \refb{gcaD}and the ``cylinder" co-ordinates \refb{cyl-NR} is given by 
\be{}
t = e^{in \t}, \quad x = i \s e^{in\s}
\ee 

\subsection{The ultra-relativistic limit in $D=2$}
The fact that one can obtain the 2d GCA by looking at the non-relativistic contraction of two copies of the Virasoro algebra is not surprising. What is surprising, however, is that there exists another contraction of the Virasoro algebra which generates the 2d GCA:
\be{ur-lim}
\L_n - \bL_{-n} = L_n, \quad \L_n + \bL_{-n} = \frac{1}{\e} M_n
\ee
Looking at the expression for the generators on the cylinder, it is easy to convince oneself that the correct space-time contraction that one is looking at now is $\s \to \s, \, \t \to \e \t$. This leads to 
\be{cyl-UR}
L_n = e^{in\s} \left( \p_\s + i n \t \p_\t \right), \quad M_n = e^{in\s} \p_\t
\ee
which is exactly \refb{cyl-NR} with $\s \leftrightarrow \t$. The interpretation of this limit is reversed from the previous one. This is the limit where the speed of light tends to zero and is hence the ultra-relativistic limit of the conformal algebra. The mapping from the 2D CFT also changes the interpretation of the central terms in this limit. Now the central terms are given by 
\be{c-ur}
c_L = c - \bar{c}, \quad c_M = \e (c + \bar{c}) \ee
which is to be contrasted to \refb{c-nr}. 

The strange co-incidence of the non-relativistic and ultra-relativistic limit displaying the same symmetry algebra is a feature unique to two dimensions where the space and time directions can be swapped and the symmetry is (at least locally) blind to this exchange. Thought about in terms of Euclidean 2D field theories, it is clear in hindsight that the theory should not distinguish between what is time and what is space and hence a contraction in one is equivalent to a contraction in the other direction. In dimensions higher than two, an ultra-relativistic and a non-relativistic contraction are no longer the same, if we demand that all the spatial directions scale in the same way. This is simply because in the ultra-relativistic case, there is only one contracted direction while in a $(D+1)$ dimensional theory, $D$ of the spatial dimensions would scale in the non-relativistic limit{\footnote{One can obviously devise skewed limits where only one spatial dimension is non-relativistic and the rest (D-1) are not. Such limits would again yield algebras isomorphic to the ones obtained by ultra-relativistic contractions.}}.

Very surprisingly, this ultra-relativistic contraction of the 2D conformal algebra has been of use in understanding aspects of holography in 3D Minkowski space times. The asymptotic symmetry algebra of flat space times in three and four bulk dimensions has been known to be the Bondi-Metzner-Sachs (BMS) algebra \cite{Bondi:1962px}. In 3D, the BMS$_3$ algebra \cite{Barnich:2006av} is actually isomorphic to the 2d GCA \cite{Bagchi:2010eg}. The limit from AdS to flat-space can be shown to induce an ultra-relativistic contraction on the 2D dual field theory converting the relativistic conformal symmetries to Galilean Conformal symmetries \cite{Bagchi:2012cy}. (In higher dimensions, as explained above, the contraction yields symmetries which are not isomorphic to the GCA.) So the dual theory to 3D Minkowski space times is a 2D GCFT. This proposal, which at times goes under the name of the BMS/GCA correspondence, has been successfully employed to understand several aspects of bulk physics in flat space \cite{Bagchi:2012cy} -- \cite{Bagchi:2014iea}.   

\subsection{Representations of 2D GCFT}
We would be interested in building the representation theory of GCFTs in close analogy with regular CFTs and hence would concentrate on the highest weigh representations \cite{Bagchi:2009pe, Bagchi:2009ca}. The states would be labeled by their eigenvalues under the dilatation operator $L_0$ and additionally under the boost $M_0$, since $[L_0, M_0]=0$: 
\be{}
L_0 |h_L, h_M \> = h_L |h_L, h_M \>, \quad M_0 |h_L, h_M \> = h_M |h_L, h_M \>
\ee
Using the two different limits, we see that the GCFT weights can be expressed in terms of the Virasoro weights $h, \h$ (where the Virasoro states are labeled as $\L_0  |h, \h\>= h |h, \h\>, \, \bL_0 |h, \h\>= \h |h, \h\>$) as:
\be{}
\mbox{NR limit:} \quad h+\h = h_L, \, h- \h = \frac{1}{\e} h_M; \quad \mbox{UR limit:} \quad h-\h = h_L, \, h+\h = \frac{1}{\e} h_M
\ee

In analogy with CFTs, there exists a notion of primary states which are annihilated by all positive modes $L_n, M_n$. For this, we assume that $h_L$ is bounded from below. We will also assume $h_M$ is bounded for later purposes. So,
\be{}
\mbox{Primary states:} \, |h_L, h_M \>_p  \quad \Rightarrow \quad L_n |h_L, h_M \>_p =  M_n |h_L, h_M \>_p = 0, \quad \forall \, n > 0
\ee
The GCA modules are built on these primary states by the action of the raising operators $L_{-n}, M_{-n}$. 

It is interesting to note here that the highest weight representations of the Virasoro algebra descend to the highest weight representations of the GCA in the NR limit but since there is a mixing of raising and lowering Virasoro operators in \refb{ur-lim}, the UR limit does not work in the same way, i.e. one cannot obtain the highest weight representations of the GCA from the Virasoro highest weights by the UR limit. 

However, it is important to point out that the highest weights of the GCA can be defined independent of any limit and the representations would just depend on intrinsically GCFT quantities like the weights $h_L, h_M$ and the central terms $c_L, c_M$.  

\subsection{Energy-momentum tensor for 2D GCFT}
One of the central objects in a 2D CFT and indeed of any quantum field theory is the Energy-Momentum tensor. It is important to define a similar object in the 2D GCFT. The holomorphic and anti-holomorphic EM tensor for a 2D CFT is defined in terms of the modes of the Viraroso algebra on the complex plane as:
\be{} 
T(z) = \sum_n  \L_n z^{-n-2}, \quad \bar{T}(\z) = \sum_n \bL_n  \z^{-n-2}
\ee
In the non-relativistic limit, one can define the combinations of the above which remain finite \cite{Bagchi:2010vw}
\be{}
T^{\mbox{\tiny{P}}}_1 (x, t) = T + \bar{T} = \sum_n  \left(L_n + (n+2)  M_n \frac{x}{t} \right) t^{-n-2}, \quad T^{\mbox{\tiny{P}}}_2 (x, t) = \e ( T - \bar{T}) = \sum_n  M_n t^{-n-2}
\ee
where the $P$ superscript denotes that the expressions are the ones for the plane representation. It is instructive to also look at the expressions of these on the cylinder. We need to keep in mind that the expressions for the EM tensor on the cylinder of a 2D CFT involves a central term arising from the Schawrzian derivative which reflects the fact that the EM tensor is not a primary field. Taking similar limits on the transformed 2D CFT EM tensor, we obtain:
\bea{Tnr}
&&T^{\mbox{\tiny{C}}}_1(\s, \t) = T^{\mbox{\tiny{C}}} + \bar{T}^{\mbox{\tiny{C}}} = \sum_n  \left(L_n - i n \s  M_n \right) e^{-in \t} + \frac{c_L}{12}, \\
&&T^{\mbox{\tiny{C}}}_2(\s, \t) = \e \left(T^{\mbox{\tiny{C}}} - \bar{T}^{\mbox{\tiny{C}}}\right) = \sum_n M_n e^{-in \t} + \frac{c_M}{12}
\eea
In the ultra-relativistic case, if we look at the cylinder again, we need to use different linear combinations to obtain finite answers in the limit:
\bea{Tur}
&&T^{\mbox{\tiny{C}}}_1(\s, \t) = T^{\mbox{\tiny{C}}} - \bar{T}^{\mbox{\tiny{C}}} = \sum_n  \left(L_n - i n \t  M_n \right) e^{-in \s} + \frac{c_L}{12}, \label{Tur1}\\
&&T^{\mbox{\tiny{C}}}_2(\s, \t) = \e \left(T^{\mbox{\tiny{C}}} + \bar{T}^{\mbox{\tiny{C}}}\right) = \sum_n M_n e^{-in \s} + \frac{c_M}{12} \label{Tur2}
\eea
So we see that \refb{Tnr} and \refb{Tur} are related just by the exchange $\s \leftrightarrow \t$. 

Thus, in a generic 2D GCFT with a contracted direction $\xi_1$ and a non-contracted direction $\xi_2$, which is labeled by its central terms $c_L$ and $c_M$, the generators on (the degenerate version of) the cylinder are given by 
\be{}
L_n = e^{in\xi_2} \left( \p_2 + i n \xi_1 \p_1 \right), \quad M_n = e^{in\xi_2} \p_1
\ee
and the EM tensor is of the form:
\be{T}
T^{\mbox{\tiny{C}}}_1(\xi_1, \xi_2) = \sum_n  \left(L_n + i n \xi_1  M_n \right) e^{in \xi_2} + \frac{c_L}{12}, \quad T^{\mbox{\tiny{C}}}_2(\xi_1, \xi_2) = \sum_n M_n e^{in \xi_2} + \frac{c_M}{12}
\ee

\subsection{Truncation of symmetry algebra}

Before we conclude our discussions of GCFTs, there is another important thing that is going to be relevant to us in our analysis of tensionless strings. Let us briefly mention this. Like in ordinary CFTs, in GCFTs there are null states (states that are orthogonal to all states in the Hilbert space including themselves) which have to be removed from the physical spectrum. The analysis in \cite{Bagchi:2009pe} treats this aspect of GCFTs in some detail. The most interesting feature that arises out of this is that there exists a sector in a GCFT where there is a truncation of the symmetry algebra. If we are dealing with a GCFT with $c_M=0$ and concentrate on a sector of the theory that has states with $h_M=0$, then through an analysis of null vectors, it can be shown that the representations of the GCA naturally reduce to the representations of the Virasoro algebra. This feature has been exploited in \cite{Bagchi:2012yk} to construct a theory of gravity in flat-spacetimes that is chiral. 

Let us revisit this analysis briefly. To find a null state at a given level, let us consider a most general state for that level by considering a combination of $L_{-n}$ and $M_{-n}$ $(n>0)$ acting on the GCA primary state $|h_L,h_M\rangle$. Then we impose the condition that $L_{n}$ and $M_{n}$ for any $n$ anhilate this state. We only require to check upto $n=2$ to obtain the conditions that fix the relative coefficients in the linear combination as well as give a relation between $h_L$, $h_M$ and the central charges $c_L$,$c_M$.

At the level $n=1$ we can only have a combination of the states $L_{-1}|h_L,h_M\rangle$ and $M_{-1}|h_L,h_M\rangle$. There 
exists only one null state for $h_M=0$. For level two we have to write the general state $|\chi\rangle$ as given below, and impose that $L_2$ and $M_2$ anhilate this state.
\be{} 
|\chi\rangle=(a_1 L_{-2}+a_2 L_{-1}^2+b_1 L_{-1}M_{-1}+d_1 M_{-1}^2+d_2 M_{-2})|h_L,h_M\rangle 
\ee
After using the GCA Algebra as in \refb{gca2d} we obtain six conditions:
\bea{} && 3a_1+2(h_L+1)a_2+2h_M b_1 =0, \quad (4h_L+6c_L)a_1+6h_L a_2+6h_M b_1+(6c_M+4h_M)d_2 =0 \nonumber \\
&& 2(h_L+1)b_1+4h_M d_1+3 d_2=0, \quad (4h_M+6 c_M)a_1=0, \quad 3a_1+2a_2+2h_M b_1=0, \quad h_M a_2 =0 \nonumber
\eea 
We can analyse the above equations by considering various cases and fixing certain values for some parameters. For $c_M\neq 0$ we can obtain certain null states at a general level $k$, with conditions relating $k$, $h_M$ and $c_M$. 

In section 5.3 we had argued the possiblilty of $c_M$ being zero in the tensionless string theory. So we will restrict our analysis following \cite{Bagchi:2009pe} on the sector of the theory having $c_M=0$. In this case we have non-trivial null states only for $h_M=0$. For $h_L=-\frac{3}{2}c_L$, we will get $a_1,a_2 \neq 0$, but only if $h_L=c_L=0$, which is a trivial case. For $h_L\neq-\frac{3}{2}c_L$ (the only contraint on the value of $c_L$ and $h_L$), we find that $a_1=a_2=0$, $d_2=-\frac{2(h_L+1)}{3}b_1$ and $d_1$ cannot be determined. This gives us the two null states as:
\be{} 
|\chi^{(1)}\rangle=M_{-1}^2|h_L,0\rangle,\ \ \ \ |\chi^{(2)}\rangle=\left[L_{-1}M_{-1}-\frac{2(h_L+1)}{3}M_{-2}\right]|h_L,0\rangle.
\ee
It is easily seen that $|\chi^{(1)}\rangle$ is a descendant of the level one null state $M_{-1}|h_L,0\rangle$. So is the state $L_{-1}M_{-1}|h_L,0\rangle$. Thus we see that if we set the level $one$ null state to zero along with the descendants then at level $two$ the new null state is simply $M_{-2}|h_L,0\rangle$. We can then set this null state and its descendants to zero. In this manner if we proceed we will obtain a new null state $M_{-k}|h_L,0\rangle$ at the $k^{th}$ level setting all the other lower level null states and their descendants to zero. Thus, if all the null states are truncated in a consistent manner, we are left with only states that are given by the Virasoro descendants of the primary. As seen above for level $two$, and is true for any general level, we only require $h_M=c_M=0$ with no generic condition on $c_L,h_L$. Then the Hilbert space of the GCA can be truncated to the Hilbert space of the Virasoro.  

In the Virasoro module, by the usual analysis of 2d CFT null vectors, one can put unitarity constraints on the values of central charge $c_L$ and weight $h_L$. In conclusion, we can have uni-tary representations of the GCA with $c_M = h_M = 0$. We call this the chiral truncation of the GCA.

\newpage

\section{The worldsheet NR Limit} \label{ApB}
In our review of GCFTs, we had remarked that there are two different contractions of the Virasoro algebra which result in the GCA. So far in our discussions, we have focused on the ultra-relativistic limit on the worldsheet. We have also mentioned the non-relativistic limit on the string world sheet briefly in Sec.~\ref{NRws} where we showed that this limit arises out of a different choice of gauge in the tensionless string. In this appendix, we attempt a construction of mode expansions similar to that of the other limit dealt with in details in the main paper. 

\subsection{Residual symmetries}
We have already seen that the gauge that needs to be fixed in this limit is 
\be{}
V^\a = (0, v)
\ee
and the residual symmetry is the 2D GCA now with generators 
\be{gen} 
L_n = ie^{in\t}( \p_\t+ in\s \p_\s), \quad M_n =  i  e^{in\t} \p_\s. 
\ee
These generators are the same as \refb{cyl-NR} which arise from the non-relativistic contraction of the Virasoro algebra \refb{nr-lim}. 

\subsection{Equation of motion and mode expansions}
The equation of motion \refb{eom} in this gauge now takes the simplified form:
\be{}
{X}^{\prime\prime \mu} =0; \quad {X'}^2=0, \quad \dot{X}\cdot X'=0. 
\ee
The mode expansion for this gauge turns out to be the following: 
\be{nr-exp}
X^{\mu}(\sigma,\tau)=x^{\mu}+\sqrt{2 c'}\A^{\mu}_0\tau+ \sqrt{2 c'}\B^{\mu}_0\s+ i\sqrt{2c'}\sum_{n\neq0}\frac{1}{n}\left(\A^\mu_n -i n \s \B^\mu_n\right) e^{-in \t}
\ee
Boundary conditions $X^\mu(\s, \t) = X^\mu (\s + 2 \pi, \t)$ now dictates that $\B^{\mu}_0 =0$. The entire analysis of constraints of Sec~\ref{TEom} and identification of this with the GCFT EM tensor follows with an exchange of $\s \leftrightarrow \t$. One now compares the constraints with the expressions of the Energy momentum tensor in the non-relativistic limit \refb{Tnr}. The oscillators $\A, \B$ follow the same algebra 
\be{}
\{ \A^{\mu}_m, \A^{\nu}_n\}_{P.B}= \{ \B^{\mu}_m, \B^{\nu}_n\}_{P.B} = 0, \quad \{ \A^{\mu}_m, \B^{\nu}_n\}_{P.B}= -2im\delta_{m+n}\eta^{\mu \nu}. 
\ee
Here the definitions of the modes of the algebra $L_n$ and $M_n$ in terms of $\A$ and $\B$ stay the same as before, viz. $L_n = \sum \A_{n+m} \B_{-m}, \ \ M_n = \sum \B_{n+m} \B_{-m}$.

\subsection{The peculiarities of the other contraction}

The interesting departure in the analysis from the previous case arises in the limit from the tensile theory. We now need to consider the limit 
\be{nr-tless}
\t \to \t, \s \to \e \s, \a' \to c'/\e 
\ee
on the mode expansion of the tensile theory \refb{t-exp}. If we continue with the identifications from the mode expansions, we get This yields \refb{nr-exp} with the following identifications:
\be{nrmap-wrong}
\A_n^\mu =  \frac{1}{\sqrt{\e}} \left( \a_n^\mu + \ta_n^\mu \right), \quad \B_n^\mu = {\sqrt{\e}} \left( \a_n^\mu - \ta_n^\mu \right)
\ee
If one tries to reproduce \refb{AB}, from here using the commutators of $\a$ and $\ta$, there is a mismatch. We clearly see that we have to introduce a relative factor of $i$ on the right hand side of the definition \refb{nrmap-wrong} to rectify the situation. Reversing the argument, we now demand 
\be{nrmap}
\A_n^\mu =  \frac{1}{\sqrt{\e}} \left( \a_n^\mu + \b_n^\mu \right), \quad \B_n^\mu = {\sqrt{\e}} \left( \a_n^\mu - \b_n^\mu \right)
\ee
Requiring \refb{AB}, then requires
\be{}
\{ \a_n^\mu, \a_m^\nu \} = -i n \delta_{n+m, 0} \ \eta^{\mu \nu}, \quad \{ \b_n^\mu, \b_m^\nu \} = i n \delta_{n+m, 0} \ \eta^{\mu \nu}
\ee
The above effectively means that $\b = i \ta$ which in term implies that in order to reproduce \refb{nr-exp}, one needs to start off with a string theory solution which is 
\be{}
X = X_L ( \t + \s) + i X_R (\t - \s)
\ee
instead of the usual $X= X_L + X_R$. This obviously solves the equation of motion of the tensile string, but there are differences with the usual solution. The mode expansion of this solution in the tensile theory is now
\be{}
X(\t, \s) = i\sqrt{2\alpha'}\sum_{n\neq0}\frac{1}{n}[\a^{\mu}_ne^{-in(\tau+\sigma)}+\b^{\mu}_ne^{-in(\tau-\sigma)}] 
\ee
Demanding reality of the solution means that there can be no zero mode in this solution. This additionally means that $A_0 =0$ which is a restriction on the solution \refb{nr-exp}. Reality of $X(\s, \t)$ also implies that the Hermiticity condition for the oscillators are
\be{}
(\a^\mu_n)^\dagger = \a^\mu_{-n}, \quad (\b^\mu_n)^\dagger = - \b^\mu_{-n}
\ee
There are obvious differences with the usual tensile theory when one attempts to quantize this solution. At this juncture, we are not very clear about what lessons this has for the tensionless theory we are looking to study. It is very clear that while the choice of gauge $V^\a = (0, v)$ locally yields the same residual gauge symmetry, the GCA as the previous case $V^\a = (v, 0)$, there are interesting differences.  

Let us quickly revisit the constraint analysis in this limit on the string worldsheet. The constraints in the tensile case are: 
\be{} 
\dot{X}^2+X^{'2}=0, \quad \dot{X} \cdot X^{\prime} =0. 
\ee 
On taking the limit $(\t, \s) \to (\t, \e \s)$ :
\be{}
\dot{X}^2+\frac{1}{\e^2} X^{'2}=0, \ \dot{X} \cdot X^{\prime} =0\quad \text{or} \quad {\epsilon^2}\dot{X}^2+X^{'2}=0, \ \dot{X} \cdot X^{\prime} =0 
\ee %L
Repeating the analysis of Sec~\ref{limit-t}, we find the tensionless constraints in terms of the tensile ones in this limit now generate the NR limit of the Virasoro algebras to the GCA \refb{} on the worldsheet:
\be{} 
L_n={\L}_n+\bar{\L}_{n} \quad M_n=\e\left[{\L}_n - \bar{\L}_{n} \right] 
\ee
Before we conclude this appendix, let us make a quick and curious observation. The rather strange complex combinations of the left and right movers that we have advocated in the construction of the NR limit on the worldsheet is reminiscent of similar complex combinations in \cite{Witten:1988zd}, where this analysis is also linked to the high energy scatterings of string theory \cite{Gross:1987kza, Gross:1987ar}. It could be worthwhile to pursue this line of argument further and link up to our present analysis of the non-relativistic limit on the worldsheet.

%\newpage

\end{document}